\documentclass[a4paper,12pt]{article}
\pdfoutput=1
\usepackage{graphicx,rotating,hyperref,slashed,amsmath,xcolor,amssymb,amsfonts}
\makeatletter
\hypersetup{colorlinks,bookmarksopen,bookmarksnumbered,
linkcolor=blus,pdfstartview=FitH,urlcolor=rossos,citecolor=verde}

\def\lsim{\mathrel{\rlap{\lower3pt\hbox{\hskip0pt$\sim$}}
   \raise1pt\hbox{$<$}}}         
\def\gsim{\mathrel{\rlap{\lower4pt\hbox{\hskip1pt$\sim$}}
   \raise1pt\hbox{$>$}}}         

 \newcommand{\MDM}{M_{\rm DM}}
 \renewcommand{\Im}{{\rm Im}\,}
\renewcommand{\Re}{{\rm Re}\,}
\newcommand{\mio}[1]{}

\newcommand{\fig}[1]{~\ref{fig:#1}}

\definecolor{Gray}{gray}{0.95}

\newcommand{\GM}{0.06}
\newcommand{\X}{S}
\newcommand{\Q}{{\cal Q}}
\newcommand{\Excess}{750\GeV}
\newcommand{\pb}{\,{\rm pb}}
\newcommand{\fb}{\,{\rm fb}}
\newcommand{\NTC}{N_{\rm TC}}

\newcommand{\sfrac}[2]{#1/#2}

\usepackage{multicol}
\usepackage{color}
\definecolor{rosso}{cmyk}{0,1,1,0.4}
\definecolor{rossos}{cmyk}{0,1,1,0.55}
\definecolor{rossoc}{cmyk}{0,1,1,0.2}
\definecolor{blu}{cmyk}{1,1,0,0.3}
\definecolor{blus}{cmyk}{1,1,0,0.6}
\definecolor{bluc}{cmyk}{1,1,0,0.1}
\definecolor{verde}{cmyk}{0.92,0,0.59,0.25}
\definecolor{verdec}{cmyk}{0.92,0,0.59,0.15}
\definecolor{verdes}{cmyk}{0.92,0,0.59,0.4}

\newcommand{\eq}[1]{~{\rm (\ref{eq:#1})}}

\newcommand{\GeV}{\,{\rm GeV}}
\newcommand{\TeV}{\,{\rm TeV}}
\newcommand{\cm}{\,{\rm cm}}
\newcommand{\Tr}{\,{\rm Tr}}

\def\circa#1{\,\raise.3ex\hbox{$#1$\kern-.75em\lower1ex\hbox{$\sim$}}\,}

\newcommand{\beq}{\begin{equation}}
\newcommand{\eeq}{\end{equation}}

\newcommand{\bea}{\begin{eqnarray}}
\newcommand{\eea}{\end{eqnarray}}
\newcommand{\be}{\begin{equation}}
\newcommand{\ee}{\end{equation}}
\font\tenrsfs=rsfs10 at 12pt
\font\sevenrsfs=rsfs7 at 10 pt
\font\fiversfs=rsfs5
\newfam\rsfsfam
\textfont\rsfsfam=\tenrsfs
\scriptfont\rsfsfam=\sevenrsfs
\scriptscriptfont\rsfsfam=\fiversfs
\def\mathscr#1{{\fam\rsfsfam\relax#1}}

\def\Lag{\mathscr{L}}

\def\circa#1{\,\raise.3ex\hbox{$#1$\kern-.75em\lower1ex\hbox{$\sim$}}\,}
\makeatletter

\def\hhref#1{\href{http://arxiv.org/abs/#1}{arXiv:#1}} 

\def\hhref#1{\href{http://arxiv.org/abs/#1}{arXiv:#1}} 
 
\def\art{\@ifnextchar[{\eart}{\oart}}
\def\eart[#1]#2#3#4#5#6{{\rm #2}, {\em #3 \bf #4} {\rm (#6) #5} ({\em #1})}

\def\article{\@ifnextchar[{\earticle}{\oarticle}}
\def\oarticle#1#2#3#4#5#6{{\rm #1}, {\em ``#6''}, {\rm #2 #3 (#5) #4}}
\def\earticle[#1]#2#3#4#5#6#7{{\rm #2}, {\em ``#7''}, {\rm #3 #4 (#6) #5}  [\hhref{#1}]}
\def\hepart[#1]#2{{\rm #2, \em#1}}
\def\heparticle[#1]#2#3{#2, {\em ``#3''} [\hhref{#1}]}

\newcounter{alphaequation}[equation]
\def\thealphaequation{\theequation\hbox to
0.6em{\hfil\alph{alphaequation}\hfil}}
\def\eqnsystem#1{
\def\@eqnnum{{\rm (\thealphaequation)}}
\def\@@eqncr{\let\@tempa\relax \ifcase\@eqcnt \def\@tempa{& & &} \or
  \def\@tempa{& &}\or \def\@tempa{&}\fi\@tempa
  \if@eqnsw\@eqnnum\refstepcounter{alphaequation}\fi
\global\@eqnswtrue\global\@eqcnt=0\cr}
\refstepcounter{equation} \let\@currentlabel\theequation \def\@tempb{#1}
\ifx\@tempb\empty\else\label{#1}\fi
\refstepcounter{alphaequation}
\let\@currentlabel\thealphaequation
\global\@eqnswtrue\global\@eqcnt=0 \tabskip\@centering\let\\=\@eqncr
$$\halign to \displaywidth\bgroup \@eqnsel\hskip\@centering
$\displaystyle\tabskip\z@{##}$&\global\@eqcnt\@ne
\hskip2\arraycolsep\hfil${##}$\hfil& \global\@eqcnt\tw@\hskip2\arraycolsep
$\displaystyle\tabskip\z@{##}$\hfil
\tabskip\@centering&\llap{##}\tabskip\z@\cr}
\def\endeqnsystem{\@@eqncr\egroup$$\global\@ignoretrue} \makeatother

\newcommand{\SU}{\,{\rm SU}}
\newcommand{\SO}{\,{\rm SO}}
\newcommand{\U}{\,{\rm U}}
\begin{document}

\centerline{CERN-PH-TH/2015-302\hfill IFUP-TH/2015}
\bigskip
\bigskip

\begin{center}
{\LARGE \bf \color{rossos} What is the $\gamma\gamma$ resonance at 750 GeV?}\\[1cm]

\bigskip\bigskip

{\large\bf 
Roberto Franceschini$^{a}$,
Gian F. Giudice$^{a}$,\\[2mm]
Jernej F. Kamenik$^{a,b,c}$,
Matthew McCullough$^a$, 
Alex Pomarol$^{a,d}$,\\[2mm]
Riccardo Rattazzi$^{e}$,
Michele Redi$^{f}$,
Francesco Riva$^a$,\\[2mm]
Alessandro Strumia$^{a,g}$,
Riccardo Torre$^{e}$
}  
\\[5mm]

\bigskip

{\it $^a$ CERN, Theory Division, Geneva, Switzerland}\\[1mm]
{\it $^{b}$ Jo\v zef Stefan Institute, Jamova 39, 1000 Ljubljana, Slovenia}\\[1mm]
{\it $^{c}$ Faculty of Mathematics and Physics, University of Ljubljana, Jadranska 19, \\1000 Ljubljana, Slovenia}\\[1mm]
{\it $^{d}$ Dept. de F\'isica and IFAE-BIST, Universitat Aut\`onoma de Barcelona, \\08193 Bellaterra, Barcelona, Spain}\\[1mm]
{\it$^{e}$ Institut de Th\'eorie des Ph\'enom\`enes Physiques, EPFL,  CH--1015 Lausanne, Switzerland}\\[1mm]
{\it $^f$ INFN, Sezione di Firenze, Via G. Sansone, 1, I-50019 Sesto Fiorentino, Italy}\\[1mm]
{\it $^g$ Dipartimento di Fisica dell'Universit{\`a} di Pisa and INFN, Italy}


\bigskip

\vspace{1cm}
{\large\bf\color{blus} Abstract}
\begin{quote}\large
Run 2 LHC data show hints of a new resonance in the diphoton distribution at an invariant mass of 750 GeV. We analyse the data in terms of a new boson, extracting information on its properties and exploring theoretical interpretations.  Scenarios covered include a narrow resonance and, as preliminary indications suggest, a wider resonance. If the width indications persist, the new particle is likely to belong to a strongly-interacting sector. We also show how compatibility between Run 1 and Run 2 data is improved by postulating the existence of an additional heavy particle, whose decays are possibly related to dark matter.
\end{quote}

\thispagestyle{empty}
\end{center}

\setcounter{page}{1}
\setcounter{footnote}{0}

\newpage
\tableofcontents

\section{Introduction}
The ATLAS and CMS collaborations have recently presented the first data obtained at the LHC Run 2 with $pp$ collisions at
energy $\sqrt{s}=13\TeV$~\cite{data}.

The ATLAS collaboration has 3.2 fb$^{-1}$ of data and claims an excess in 
the distribution of events containing two photons, at the diphoton invariant mass $M\approx \Excess$
with $3.9\sigma$ statistical significance ($2.3\sigma$ after including the look-elsewhere effect).
The ATLAS excess consists of about 14 events (with selection efficiency 0.4) which appear in at least two energy bins,
suggesting a best-fit width of about 45~GeV ($\Gamma/M \approx \GM$),
although the very existence of this feature is uncertain.

The result is partially corroborated by the CMS collaboration with integrated luminosity of 2.6 fb$^{-1}$, which has reported
a mild excess of about 10 $\gamma\gamma$ events, peaked at 760 GeV.
The best fit has a narrow width 
and a local statistical significance of $2.6\sigma$.
Assuming a large width $\Gamma/M \approx \GM$, the significance decreases to $2.0\sigma$,
corresponding to a cross section of about 6 fb.

The anomalous events are not accompanied by significant missing energy, nor leptons or jets.
No resonances at invariant mass $\Excess$ are seen in the new data in $ZZ$, $\ell^+\ell^-$, or $jj$ events.
No $\gamma\gamma$ resonances were seen in Run 1 data  at $\sqrt{s}=8\TeV$, altough 
both CMS and ATLAS data showed a mild upward fluctuation at  $m_{\gamma\gamma}=\Excess$.
The excess in the cross sections in the $m_{\gamma\gamma}$ interval, roughly corresponding to the claimed width,
can be estimated as:
\beq\label{eq:sigma813}
\sigma(pp\to\gamma\gamma) \approx 
\left\{\begin{array}{lll}
(0.5\pm0.6)\fb~&\hbox{CMS~\cite{CMSgg8}}&\sqrt{s}=8\TeV,\\
(0.4\pm0.8)\fb~&\hbox{ATLAS~\cite{ATLASgg8}}&\sqrt{s}=8\TeV,\\
(6\pm3)\fb~&\hbox{CMS~\cite{data}}&\sqrt{s}=13\TeV,\\
(10\pm3)\fb~&\hbox{ATLAS~\cite{data}}&\sqrt{s}=13\TeV.
\end{array}\right.
\eeq
The data at $\sqrt{s}=8$ and 13 TeV are compatible at $2\sigma$ if the signal cross section grows by at least a factor of 5.

While the answer to the question in the title could just be ``a statistical fluctuation'',
it is interesting to try to interpret the result as a manifestation of new physics.
In section~\ref{pheno} we assume that the signal is due to a new resonance and determine the required partial widths, relating them to an effective description in terms of non-renormalizable operators.
In section~\ref{perturbative} we present weakly-coupled renormalizable models that realise the necessary properties of the resonance. 
The total signal rate can be reproduced in simple models, while rather special ingredients are needed to reproduce also the relatively large width. An alternative explanation of the
apparently large width could come from a multiplet of narrow resonances with mass difference comparable to $\Gamma$. 
In section~\ref{nonpert} we interpret the signal in the context of strongly-interacting new physics. 
Modelling the resonance as a composite state allows for a natural explanation of the large width, as well as the partial width in the $\gamma\gamma$ channel.
In section~\ref{DM} we consider decays into Dark Matter.
In section~\ref{heavier} we discuss the compatibility between data at $\sqrt{s}=8$ and $13\TeV$ and propose a different approach to explain the absence of signals in Run 1. We speculate on the existence of a new particle, too heavy to have a significant production rate at $\sqrt{s}=8$~TeV, but much more accessible at 13~TeV. This particle decays into the $\Excess$ resonance accompanied either by invisible particles, possibly related to dark matter, or to undetected soft radiation.
Conclusions are presented in section~\ref{section}.

\section{Phenomenological analysis}\label{pheno}
We start by interpreting the excess as the resonant process $pp\to\X\to \gamma\gamma$ where $\X$ is a new uncoloured
boson with mass $M$, spin $J$, and width $\Gamma$,
coupled to partons in the proton.
The signal cross section at proton centre-of-mass energy $\sqrt{s}$ (= 8 or 13 TeV) is
\beq \sigma(pp\to \X\to \gamma\gamma) =\frac{2J+1}{M\Gamma s} \bigg[
\sum_\wp C_{\wp\bar \wp} \Gamma(\X\to \wp\bar \wp)\bigg]\Gamma(\X\to \gamma\gamma) \, ,
\label{eq:sigmasig}
 \eeq
where the relevant $\X$  decay widths are evaluated at leading order in QCD. 
The sum is over all partons $\wp=\{g,b,c,s,u,d,\gamma\}$.
The $2J+1$ factor could be reabsorbed by redefining the widths as summed over all $\X$ polarisations, rather than averaging over them. The decay into two photons implies that the two relevant cases are $J=0,2$.
As far as eq.~\eqref{eq:sigmasig} is concerned, without loss of generality, we can focus on a spin-0 resonance. The dimensionless partonic integrals are
\begin{eqnsystem}{sys:C}
C_{gg} &=& \frac{\pi^2 }{8} \int_{M^2/s}^1 \frac{dx}{x} g(x) g(\frac{M^2}{sx}),
\\
C_{\gamma\gamma} &=& 8\pi^2 \int_{M^2/s}^1 \frac{dx}{x} \gamma(x) \gamma(\frac{M^2}{sx}),\\
C_{q\bar q}&=&\frac{4\pi^2}{9}\int_{M^2/s}^1 \frac{dx}{x} \bigg[q(x) \bar{q}(\frac{M^2}{sx})+\bar q(x) q(\frac{M^2}{sx})\bigg].
\end{eqnsystem}
Their numerical values, computed for a resonance at $M=\Excess$ using the MSTW2008NLO~\cite{pdfs} set of pdfs evaluated at the scale $\mu=M$, are:
\beq\begin{array}{c|ccccccc}
\sqrt{s} & C_{b\bar b} & C_{c\bar c} & C_{s\bar s}  & C_{d\bar d}& C_{u\bar u} & C_{gg} & C_{\gamma\gamma}\\ \hline
8\TeV & 1.07 & 2.7 & 7.2 & 89  & 158& 174 & 11 \\
13\TeV  & 15.3 & 36 & 83 & 627 & 1054& 2137 & 54
\end{array}\,,
\label{eq:line}\eeq
where $C_{\gamma\gamma}$ has a $100\%$ uncertainty if extracted purely from data without relying on theory.
On the other hand, the values of $C_{\gamma\gamma}$ are reliably  extracted from theory, assuming that 
quark splittings into photons dominate the photon pdf.
Thus, the gain factors $r=\sigma_{13\TeV}/\sigma_{8\TeV}=[C_{\wp\wp}/s]_{13\TeV}/[C_{\wp\wp}/s]_{8\TeV}$ from 8 to 13~TeV are
\beq\begin{array}{ccccccc}
 r_{b\bar b} & r_{c\bar c} & r_{s\bar s}  & r_{d\bar d}& r_{u\bar u} & r_{gg} & r_{\gamma\gamma}\\ \hline
5.4 & 5.1 & 4.3 & 2.7  & 2.5 & 4.7 & 1.9
\end{array}\,.
\label{eq:reline}\eeq
Higher order QCD corrections (not included here) can modify the numbers in eq.\eq{line} by $K$ factors of order unity. Typical values at NLO are $K_{gg} =1.5$ and $K_{q\bar q}=1.2$ (c.f.~\cite{nlo}). These corrections depend on the specific channel  but
 negligibly depend on $\sqrt{s}$ because we are considering a resonant process that always occurs at the same centre-of-mass parton energy. Hence, they roughly cancel out in the gain factors $r$.

We will focus mostly on $gg$ and $b\bar b$ induced processes, which represent the extreme cases as they give the minimum and maximum value of $C$, and also lead to a large gain in parton luminosity going from 8 to 13 TeV, as needed to fit the data.
On the other hand, $S$ production from $\gamma\gamma$ (see also \cite{gamma}) is disfavoured
by the small value of $r_{\gamma\gamma}$, which has a small uncertainty,
because partonic photons are dominantly emitted from $u$ quarks, and their
pdf evolution is under good theoretical control.

\begin{figure}[t]
\begin{center}
$$\includegraphics[width=0.9\textwidth]{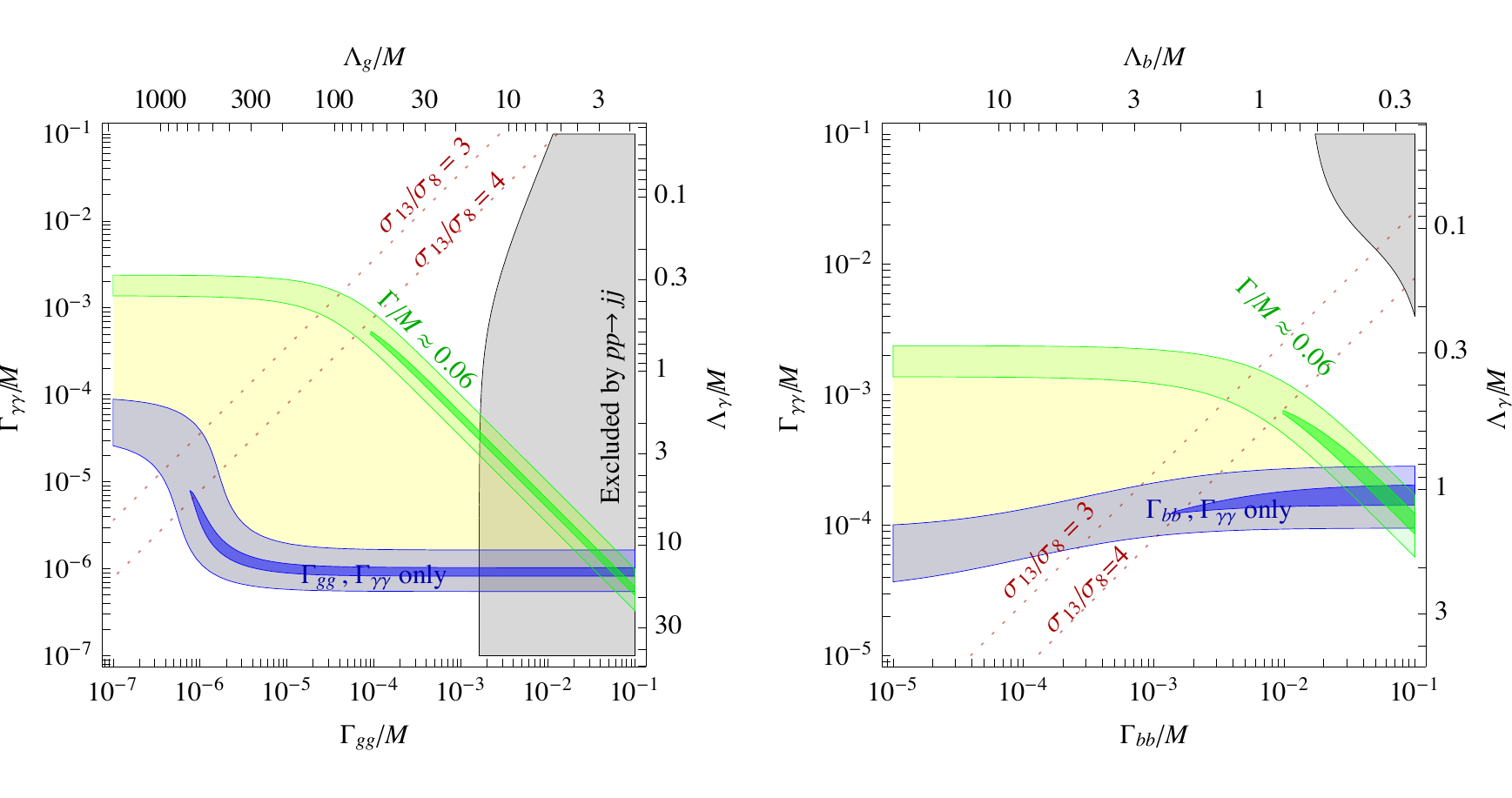}
$$
\caption{\em 
{\bf  Left (a):}
The yellow region describes the range of $\Gamma(\X\to gg)/M$  and $\Gamma(\X\to \gamma\gamma)/M$  in which the diphoton rate can be fitted as $gg\to \X\to\gamma\gamma$.
Its upper boundary is the green band (at $1\sigma$ and $2\sigma$) in which the total width is $\Gamma/M\approx \GM$, as suggested by data.
Its lower boundary is the blue band, which assumes a minimal total width $\Gamma = \Gamma(\X\to gg)+\Gamma(\X\to \gamma\gamma)$.
The grey region is excluded by searches for dijet resonances at Run 1
and is conservatively computed assuming $\Gamma = \Gamma_{gg} +\Gamma_{\gamma \gamma}$.
The upper and right axes show the values of the operator coefficients defined in eq.\eq{ops}.
The dotted lines show iso-curves of the ratio between production cross-sections at $13\TeV$ and $8\TeV$.
{\bf  Right (b):} The analogous plot, assuming that the resonant production is initiated by $b\bar b$.
\label{fig:figGG}}
\end{center}
\end{figure}


\subsection{An $s$-channel resonance coupled to gluons and photons}
Let us first consider the case in which a spin-0 resonance is produced from gluon fusion and decays into two photons. 
When production from $\gamma\gamma$ partons can be neglected with respect to production from $gg$,
the claimed signal rate is reproduced for 
\beq {\rm BR}(\X\to \gamma\gamma)\, {\rm BR}(\X\to gg)\approx  1.1\times 10^{-6} \,  \frac{M}{\Gamma} \approx 1.8\times 10^{-5},
\label{eq:BR}\eeq
or, equivalently,
\beq \frac{\Gamma_{\gamma\gamma}}{M} \, \frac{\Gamma_{gg}}{M}\approx  1.1\times 10^{-6}\, \frac{\Gamma}{M} \approx 6\times 10^{-8},
\label{eq:GM}\eeq
where $\Gamma_{\gamma \gamma } \equiv \Gamma(\X\to \gamma\gamma)$ and $\Gamma_{gg } \equiv \Gamma(\X\to gg)$.
The first set of equalities in eqs.~(\ref{eq:BR})--(\ref{eq:GM}) follows from the request $\sigma(pp\to\gamma\gamma)\approx 8\fb$ at $\sqrt{s}=13\TeV$, while the second one uses the additional information on the total width, $\Gamma / M\approx \GM$.

Figure\fig{figGG}a visualises the region of $\Gamma_{\gamma \gamma }$ and $\Gamma_{gg}$ in which the observed excess can be explained. The diphoton rate implies that the acceptable region must lie above the blue band, which is obtained by assuming no extra decay channels ($\Gamma = \Gamma_{gg} +\Gamma_{\gamma \gamma}$). Note that the blue band is essentially straight when $\Gamma_{gg} \gg \Gamma_{\gamma \gamma}$. This is because, in this limit, the total width is $\Gamma \approx \Gamma_{gg}$, and eq.\eq{GM} simplifies into $\Gamma_{\gamma \gamma}/M\approx  1.1\times 10^{-6}$, irrespectively of the value of $\Gamma$. 

\smallskip

In the opposite limit $\Gamma_{\gamma \gamma}\gg \Gamma_{gg}$, 
production from $\gamma\gamma$ partons becomes important and this is reflected in the figure
by the fact that all allowed bands become horizontal at negligible $\Gamma_{gg}$ and at
\begin{equation}\label{gagaconst}
\frac{\Gamma(S\to\gamma\gamma)}{M} =0.008 \sqrt{\frac{\Gamma}{M}} \approx 0.002
\qquad\hbox{i.e.}\qquad
{\rm BR}(\X\to \gamma\gamma) \approx 0.008  \sqrt{\frac{M}{\Gamma}}\approx 0.03.
\end{equation}
However, at the same time, Run 2 and Run 1 $\gamma\gamma$ data become incompatible such that 
a joint fit has a poor confidence level.

\medskip
 
In each point of the allowed region in fig.\fig{figGG}a above the blue band (coloured in yellow), eq.\eq{GM} determines the value of the total width. In particular, along the green band the constraint on the total width $\Gamma / M \approx \GM$ is satisfied. This is the region singled out by the ATLAS data, taken at face value. In each point of the plane in fig.\fig{figGG}a we can compute the rate of dijets induced by the decay of $\X$ back into two gluons. Searches for dijet resonances at $\sqrt{s} =8$~TeV~\cite{jjbound} rule out the grey region in the figure. Note that, for $\Gamma_{gg}>\Gamma_{\gamma \gamma}$, a resonance coupled only to gluons and photons (which corresponds to the intersection between blue and green bands) predicts a peak in $pp\to jj$ in tension with the existing experimental upper bound.

\smallskip

\begin{table}[t]
\centering{
\begin{tabular}{|c|cclc|}
\hline 
final  & \multicolumn{3}{c}{$\sigma$ at $\sqrt{s}=8\TeV$ } & implied bound on \\
state $f$ & observed & expected & ref. & $\Gamma(\X\to f)/\Gamma(\X\to \gamma\gamma)_{\rm obs}$ \\
\hline 
\hline 
$\gamma\gamma$ & $<$ 1.5 fb& $<$ 1.1 fb & \cite{CMS-PAS-HIG-14-006,Aad:2015mna} & $< 0.8 ~(r/5)$\\  
$e^+e^-  , \mu^+\mu^-$& $<$ 1.2 fb & $<$ 1.2 fb & \cite{1405.4123} &  $< 0.6~(r/5)$\\
$\tau^+\tau^-$ & $<$ 12 fb & $<$ 15 fb & \cite{1409.6064} & $< 6~(r/5)$\\
$Z\gamma$ & $<$ 11 fb & $<$ 11 fb & \cite{1407.8150} &   $< 6~(r/5)$\\
$ZZ$ & $<$ 12 fb & $<$ 20 fb & \cite{ATLAS-Collaboration:2015hb} &   $< 6~(r/5)$\\
$Zh$ & $<$ 19 fb & $<$ 28 fb & \cite{1502.04478} &   $< 10~(r/5)$\\
$hh$ & $<$ 39 fb & $<$ 42 fb  & \cite{[ATLAS-CONF-2014-005]} & $< 20~(r/5)$\\  
$W^+W^-$ & $<$ 40 fb & $<$ 70 fb& \cite{CMS-Collaboration:2015vv,ATLAS-Collaboration:2015sf} &   $< 20~(r/5)$\\\hline
$t\bar{t}$ & $<$ 450 fb & $<$ 600 fb & \cite{Chatrchyan:2013yq} & $<     300~(r/5)$\\  
invisible &  $<$ 0.8 pb &-&  \cite{1408.3583} & $< 400~(r/5)$ \\  
$b\bar b$ & \hbox{$\circa{<} 1\pb$}& \hbox{$\circa{<} 1\pb$}  &\cite{1506.08329}&  $< 500~(r/5)$ \\  
$ jj$  & $\circa{<}$ 2.5 pb &- & \cite{jjbound} & $< 1300~(r/5)$ \tabularnewline  
\hline 
\end{tabular}}
\caption{\em Upper bounds at 95\% confidence level on $pp$ cross sections at $\sqrt{s}=8\TeV$ for various final states 
produced through a resonance with $M=\Excess$ and $\Gamma/M\approx\GM$.
Assuming  that the production cross section grows as $r=\sigma_{13\TeV}/\sigma_{8\TeV}\approx 5$,
and that $\X\to \gamma\gamma$ fits the central value of the $\gamma\gamma$ anomaly,
 we show  in the last column
 the upper bounds on the partial widths in different channels. 
Similar analyses claim a bound on the $jj$ cross section which is weaker by a factor of few, and with a surprisingly large dependence
on the assumed width and shape.
%
%
%
\label{tabounds}}
\end{table}

In order to relax this constraint, it is useful to consider extra decay channels beyond $\gamma\gamma$ and $gg$. Table~\ref{tabounds} summarises the upper bounds on cross sections at 8~TeV due to an $s$-channel narrow resonance at $\Excess$,
decaying into various final states. In the last column of the table, the limit on the 8~TeV cross section is translated into a limit on the partial decay width, in units of the width into photons corresponding to the ATLAS observation. The rescaling factor  $r=\sigma_{13\TeV}/\sigma_{8\TeV}$ is about 5 for resonances produced from gluons (as well as bottom quarks), see eq.\eq{reline}.
The first entry in the table shows that rescaling the 8~TeV data constrains the diphoton peak to be at most 80\% of what observed by ATLAS. In section~\ref{heavier} we will further discuss this tension and show how it can be resolved by the production of a new particle heavier than $\X$. The other entries show that significant constraints are present in all channels. 
This holds even for a possible invisible decay of $\X$ into neutrinos or dark matter particles.
By computing the $pp\to j\X$ cross section, 
with the jet $j$ arising from initial state radiation (assuming that $pp\to \X$ comes from $gg$ partons),
and comparing it to the bounds on jets plus missing energy,
we find the constraint on the invisible width shown in the table. For the channels above the horizontal line, the constraints are strong enough that a width $\Gamma/M\approx 0.06$ cannot be reproduced without entering in conflict also with eq.~(\ref{gagaconst}). On the other hand, the weakest bound corresponds to a peak in the dijet distribution.
As long as the simplest decay channels are considered, 
the total width cannot be larger than $\Gamma \circa{<} 1500 \times \Gamma(\X\to\gamma\gamma)_{\rm obs}$. Using the ATLAS result $\Gamma /M \approx\GM$, this bound implies $\Gamma_{\gamma \gamma}/M >  4\times 10^{-5}$. This conclusion can be avoided by devising
 special final states with weaker bounds, such as many soft multi-jets.

\begin{figure}[t]
$$\includegraphics[width=0.45\textwidth]{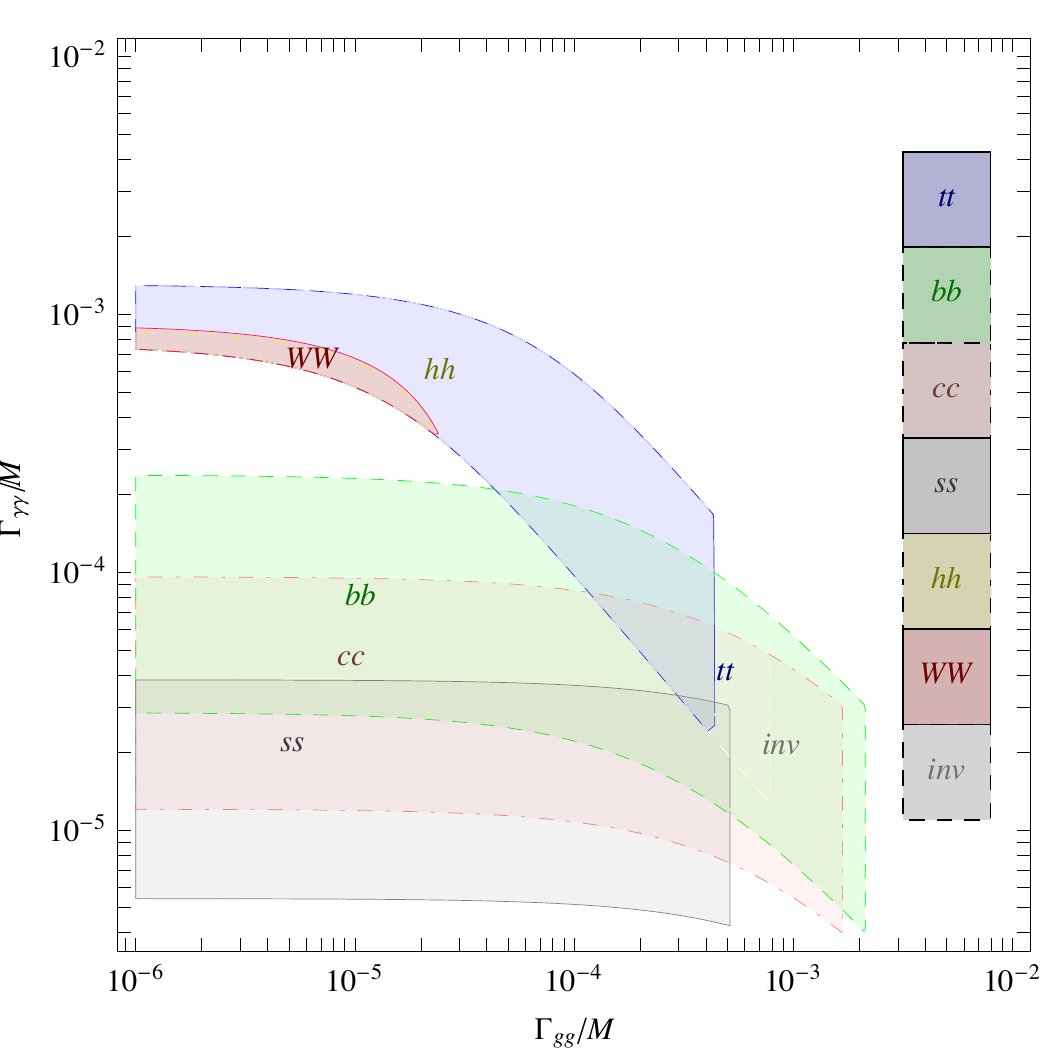}\qquad
\includegraphics[width=0.45\textwidth]{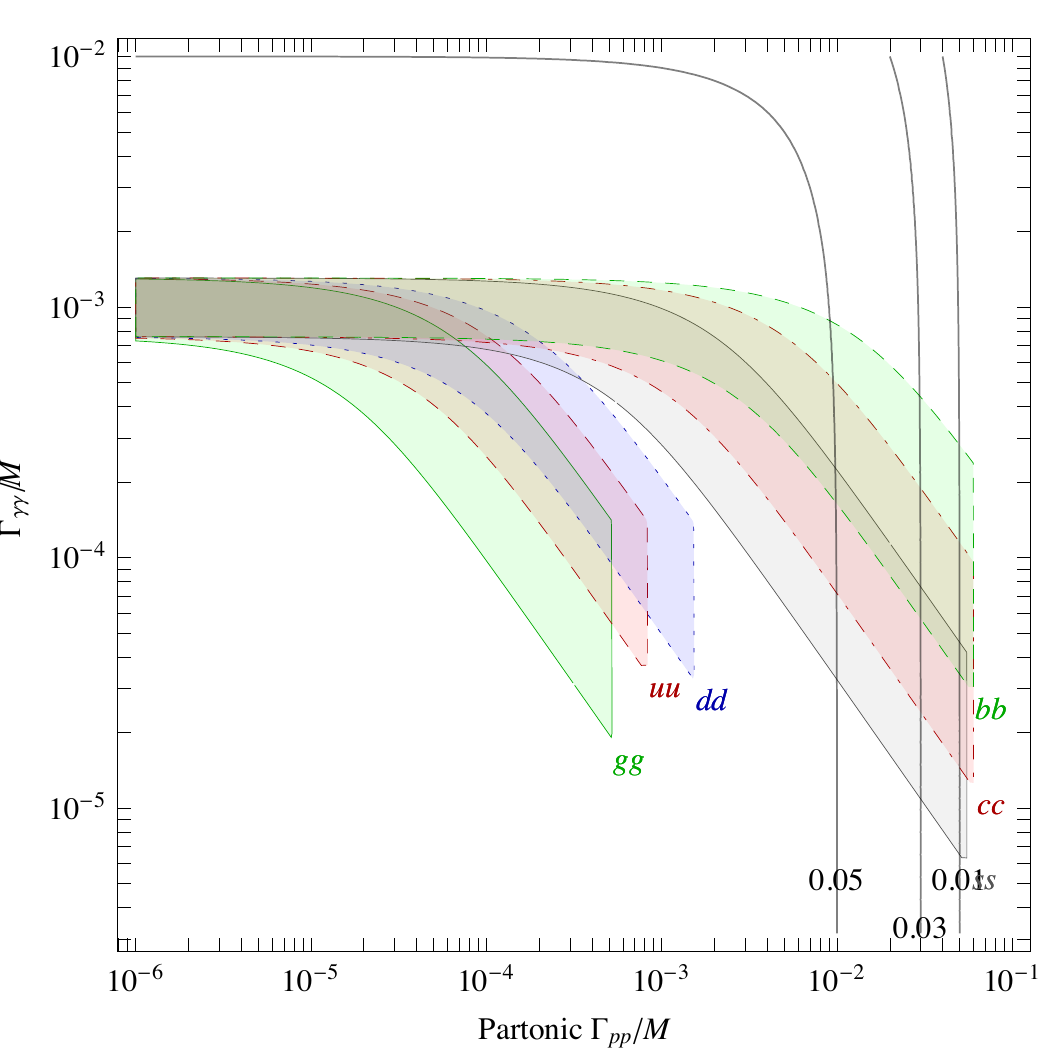}$$
  \caption{\em 
 {\bf Left panel}: regions that fit, at $3\sigma$ confidence level, the $\gamma\gamma$ rate, peak position and the large
  width (possibly suggested by ATLAS)
  assuming a resonance $\X$
  that can decay into $\gamma\gamma$, $gg$ and a third channel among those considered in the figure.
The left boundaries of the allowed regions in the diagonal band are the same for all channels, while the right boundaries differ for the individual channels and are marked by the labels.  All constraints in table~\ref{tabounds} have been taken into account. 
Decays into leptons or $ZZ$ can fit data only if $\Gamma_{\gamma\gamma}$ is large,
a possibility disfavoured by run 1 data.
{\bf Right panel}: regions that fit the diphoton excess and that satisfy all constraints
assuming that $S$ couples to a single parton $\wp$ with width $\Gamma_{\wp\wp}$, 
to photons with width $\Gamma_{\gamma\gamma}$, and to Dark Matter 
such that the total width is $\Gamma=0.06 M$.
We show contour-lines of $\Gamma_{\rm DM}/M$ and consider each parton $\wp=\{g,u,d,s,c,b\}$ in turn.
Production dominated by $u\bar u$, $d\bar d$ and especially $\gamma\gamma$ partons implies
a poor compatibility between run 1 and run 2 $\gamma\gamma$ data. 
 \label{fig:good}}
\end{figure}

The impact of the Run 1 searches for resonances on the interpretation of the ATLAS excess is visualised in the left panel of fig.\fig{good}. We assume here that $\X$ has three possible decay modes: $\gamma\gamma$, $gg$, and one of the channels listed in the figure. In each case, we show the region in which the rate and total width of the excess are explained, and all bounds from Run 1 data given in table~\ref{tabounds} are  satisfied. When the third decay channel involves quarks, the contribution to the $\X$ production cross section is included. We observe from the left panel of fig.\fig{good} that solutions are possible for all channels, although the most constrained channels ({\it e.g.} leptons) require unusually large values of $\Gamma_{\gamma \gamma}$ to explain the data.

\subsection{An $s$-channel resonance coupled to $b$ quarks and photons}\label{gammab}

We can now repeat the analysis for the case in which the resonance $\X$ is produced from bottom quark annihilation. 
In the limit $\Gamma_{b\bar b}\gg \Gamma_{\gamma\gamma}$ the signal is reproduced for
\beq \frac{\Gamma_{\gamma\gamma}}{M} \, \frac{\Gamma_{b\bar b}}{M}\approx  1.9 \times 10^{-4}\, \frac{\Gamma}{M} \approx 1.1\times 10^{-5} \, ,
\label{eq:GMb}\eeq
where, as before, the second equality follows from the further requirement $\Gamma /M \approx \GM$.
In view of the reduced $b\bar b$ parton luminosity (compared to $gg$) the range of $\Gamma_{\gamma\gamma}$ and $\Gamma_{b\bar b}$
suggested by the signal rate are now larger, and closer 
to the claimed value of the total width.
The predicted $pp\to b\bar b$ cross section is at most 0.1 pb, and therefore the search for resonances in $b\bar b$ at Run 1 does not impose a significant constraint.
The situation is illustrated in the right panel of fig.\fig{figGG}.

\subsection{Effective operators: spin 0}\label{spin0}
{Assuming the (pseudo)scalar $\X$ is the lightest state involved in the $\gamma\gamma$ excess, the previously considered  decay widths   can be described by an effective lagrangian involving  the following  operators}
\beq   {g_3^2} \X \bigg(\frac{G_{\mu\nu}^2}{{ 2\Lambda_g}} +\frac{G_{\mu\nu}\tilde G^{\mu\nu}}{2\tilde\Lambda_g}\bigg)
+ e^2\X \bigg(\frac{F_{\mu\nu}^2}{{ 2\Lambda_\gamma}} +\frac{F_{\mu\nu}\tilde F^{\mu\nu}}{2\tilde\Lambda_\gamma}\bigg)
+{  \X  \, \sum_\psi   \, \bar \psi (y_{\psi S} + i\gamma_5 \tilde y_{\psi S}) \psi   }, 
\label{eq:ops}
\eeq
where $\psi$ are the SM fermions and
$G_{\mu\nu}$ and $F_{\mu\nu}$ are the gluon and photon field strengths
and $\tilde F_{\mu\nu} = \frac12 \epsilon_{\mu\nu\alpha\beta}F^{\alpha\beta}$. To simplify the notation, we have included simultaneously the CP-even and CP-odd couplings, but it should be understood that, unless CP is badly broken, only one coupling at a time is present.
These operators give rise to
\begin{eqnsystem}{sys:Gammaop}
\Gamma(\X\to \gamma\gamma)&=& \pi \alpha^2 M \left(\frac{M^2}{\Lambda_\gamma^2} + \frac{M^2}{\tilde \Lambda_{ \gamma}^2}  \right)\,,  \\
 \Gamma(\X\to gg)&=&  8 \pi \alpha_3^2 M \left(\frac{M^2}{\Lambda_g^2} + \frac{M^2}{\tilde \Lambda_{ g}^2}  \right)\,,\\
\Gamma(\X\to \psi\bar \psi) & = & \frac{N_\psi M}{8\pi} (y_{\psi\X}^2 + \tilde y_{\psi\X}^2)
\end{eqnsystem}
where $N_\psi$ is the number of components of $\psi$ ($N_\psi=3$ for a quark).
{Focusing on a CP-even resonance, we obtain from eqs.\eq{GM} and\eq{GMb} that the experimental signal is reproduced for
\beq \frac{M}{\Lambda_\gamma}\, \frac{M}{\Lambda_g} \approx 0.14\, \sqrt{\frac{\Gamma}{M}} \approx 0.037
\qquad {\rm or}\qquad 
y_{b\X}\frac{ M}{\Lambda_\gamma}\, \approx 9\, \sqrt{\frac{\Gamma}{M}} \approx 2 \, .
\label{eq:lambdafit}
\eeq
This result is also shown in fig.\fig{figGG}, where the translation between the operator scales $\Lambda$ and the partial widths is given by the different axis labelling. For the gluon-induced process, the ``effective coupling strengths'' $ {M}/{\Lambda_\gamma}$ and $ {M}/{\Lambda_g}$ can be less than 1 corresponding to    $\Lambda>M=\Excess$, although not much larger.
On the other hand, in view of the reduced parton luminosity, $\X$ produced through $b\bar b$ pairs requires the effective couplings, $M/\Lambda_\gamma$ and $y_{b\X}$ to be  of order unity or larger. 

\medskip

The coupling of $\X$ to photons is not invariant under the SM gauge group.
Since $M$ is larger than $v=174\GeV$, it is more reasonable to assume electroweak gauge invariant operators in eq.\eq{ops}  \cite{1008.5302}. 
Assuming  $S$ is an electroweak singlet, the leading $\SU(3)_c\otimes \SU(2)_L\otimes \U(1)_Y$-invariant operators in a derivative and field expansion affecting $\X$ production and decay are then
\begin{eqnarray}  
&&
{g_3^2} \X \bigg(\frac{G_{\mu\nu}^2}{{ 2\Lambda_g}} +\frac{G_{\mu\nu}\tilde G^{\mu\nu}}{2\tilde\Lambda_g}\bigg)
+
{g_2^2} \X \bigg(\frac{W_{\mu\nu}^2}{{ 2\Lambda_W}} +\frac{W_{\mu\nu}\tilde W^{\mu\nu}}{2\tilde\Lambda_W}\bigg)
+
{g_1^2} \X \bigg(\frac{B_{\mu\nu}^2}{{ 2\Lambda_B}} +\frac{B_{\mu\nu}\tilde B^{\mu\nu}}{2\tilde\Lambda_B}\bigg)+\nonumber
\\
&&
+ \X  \left(\frac{H {\bar \psi}_L \psi_R}{\Lambda_\psi} +{\rm h.c.} \right)+
S  \frac{|D_\mu H|^2}{\Lambda_H}+
S \frac{H^\dagger D^2 H+\hbox{h.c.}}{\Lambda_S}
\label{eq:opsSU2}
\end{eqnarray}
where $H$ is the Higgs doublet, and the scale $\Lambda_\psi$ is in general complex ($\Lambda_\psi$ is real if $S$ is a scalar and  pure imaginary  if  $S$ is a pseudo-scalar).
The operators in eq.\eq{ops} are obtained with coefficients
\beq 
\frac{1}{\Lambda_\gamma} = \frac{1}{\Lambda_B}+ \frac{1}{\Lambda_W},\qquad
y_{\psi\X} = v \frac{\Re \Lambda_\psi}{|\Lambda_\psi|^2},\qquad
\tilde y_{\psi\X} = -v \frac{\Im \Lambda_\psi}{|\Lambda_\psi|^2}\,.
\eeq
Notice that we did not include the scalar potential interaction $S|H|^2$. One is easily convinced that, given $S$ has a mass, by a redefinition of $S$ such term can always be eliminated in favor of the derivative interactions already shown in eq.\eq{opsSU2}.
In the limit $M_{W,Z,h}\ll M$ one can neglect it and
the small mixing between $S$ and $h$, with angle $\tan 2\theta=\sqrt{2/\lambda}M_h \Lambda_S/(M_S^2-M_h^2)$,
finding the $S$ decay widths
\bea
\Gamma(\X\to Z\gamma)&\approx& 2\pi \alpha^2 M^3 \left[
\left(\frac{\tan\theta_{\rm W}}{\Lambda_B}- \frac{\cot\theta_{\rm W}}{\Lambda_W}\right)^2+
\left(\frac{\tan\theta_{\rm W}}{\tilde\Lambda_B}- \frac{\cot\theta_{\rm W}}{\tilde\Lambda_W}\right)^2\right]\,, \nonumber \\
\Gamma(\X\to ZZ)&\approx& \pi \alpha^2 M^3 \left[ \nonumber
\left(\frac{\tan^2\theta_{\rm W}}{\Lambda_B}+ \frac{\cot^2\theta_{\rm W}}{\Lambda_W}\right)^2+
\left(\frac{\tan^2\theta_{\rm W}}{\tilde \Lambda_B}+ \frac{\cot^2\theta_{\rm W}}{\tilde \Lambda_W}\right)^2\right]\nonumber \\
&&+\frac{M}{128\pi} \left( \frac{M}{\Lambda_H}\right)^2 \nonumber
\,,  \\
 \Gamma(\X\to W^+W^-)&\approx&  \frac{2 \pi \alpha^2 M}{\sin^4\theta_{\rm W}} \left(\frac{M^2}{\Lambda_W^2} + \frac{M^2}{\tilde \Lambda_{W}^2}  \right)+
\frac{M}{64\pi} \left( \frac{M}{\Lambda_H}  \right)^2\,,\nonumber\\
\Gamma(\X\to hh) & \approx &\frac{M}{128\pi} \left( \frac{M}{\Lambda_H} \right)^2 \,.
\label{sys:Gammaop2}
\eea
The operators in eq.\eq{opsSU2} give rise also to 3-body decays, like $\X \to ggg$ or $\X \to h b\bar b$. The latter could be especially interesting for heavy $\X$, since the 2-body decay is suppressed by $v^2/\Lambda_b^2$. However, for the range of parameters under consideration, these processes can be safely neglected.

The $\SU(2)_L$-invariant operators give rise to the following signal ratios:
\beq
\begin{array}{c|c|c|c}
\hbox{operator} &
\displaystyle\frac{\Gamma(\X\to Z\gamma)}{\Gamma(\X\to \gamma\gamma)} &
\displaystyle\frac{\Gamma(\X\to ZZ)}{\Gamma(\X\to \gamma\gamma)}&
\displaystyle\frac{\Gamma(\X\to WW)}{\Gamma(\X\to \gamma\gamma)} \\[3mm] \hline
WW~\hbox{only} & 
\displaystyle{2}/{\tan^2\theta_{\rm W}}\approx 7 &
\displaystyle {1}/{\tan^4\theta_{\rm W}}\approx 12 &
\displaystyle {2}/{\sin^4\theta_{\rm W}}\approx 40^{\phantom{1^1}}\\[1mm] 
BB~\hbox{only} &2 \tan^2\theta_{\rm W} \approx 0.6   & \tan^4\theta_{\rm W}\approx 0.08 & 0\\
\end{array}
\label{eq:ZZ}
\eeq
We see that  the decay  to $ZZ/WW$ can be suppressed if 
the hypercharge $BB$ operators are the main source of   the  decay of $S$ to photons.
Then the bounds from resonant weak gauge boson production, shown in table~\ref{tabounds}, are easily satisfied. 
{A model where the coupling of $\X$ to gauge bosons is generated by the exchange of new matter fields 
that only possess hypercharge quantum numbers will only feature $\X B_{\mu\nu}^2$ and realise this situation.}
On the other hand, the $ZZ,WW$ rates induced by  $\X W_{\mu\nu}^2$ exceed the bounds in table~\ref{tabounds} by a factor of 2.
In the presence of both operators, the bounds are satisfied for
$-0.3 < \Lambda_B/\Lambda_W,\tilde \Lambda_B/\tilde\Lambda_W < 2.4$.
Fig.\fig{GammaRatio} shows the predictions of a set of mediators, as described in the caption.

\begin{figure}[t]
\begin{center}
$$ \includegraphics[width=0.65\textwidth]{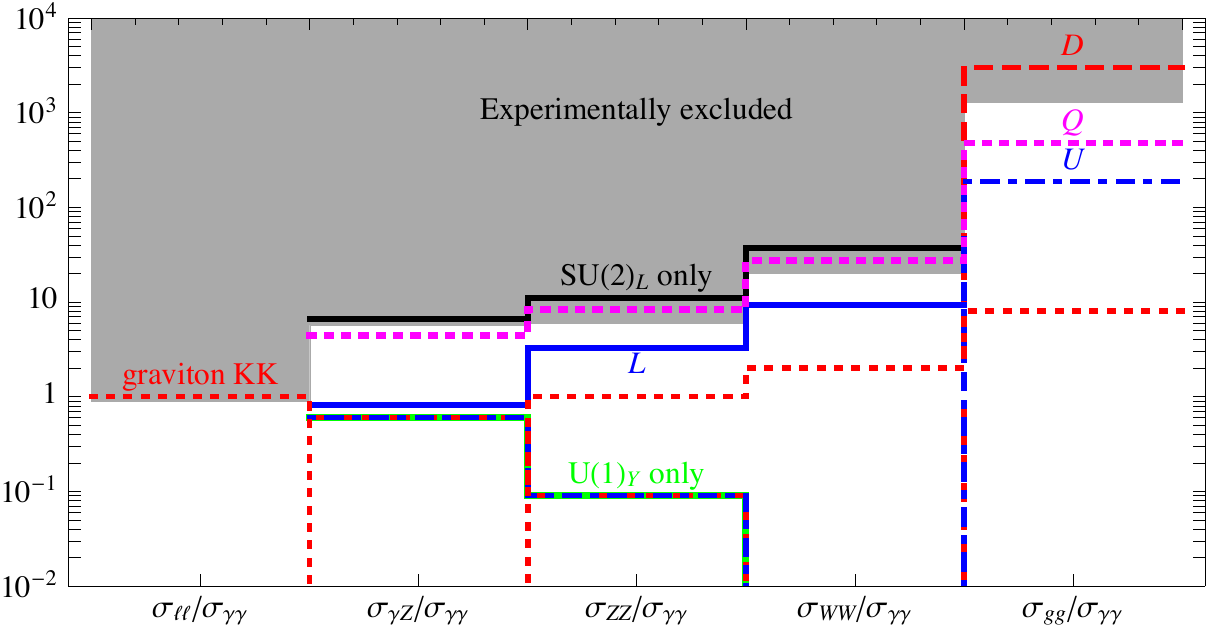}$$
\caption{\em \label{fig:GammaRatio}
Predicted cross section into various final states in units of $\sigma(pp\to\X\to\gamma\gamma)$
compared to the experimental bounds.  The models that satisfy all bounds are:
a loop of vector-like right-handed up quarks $U$  (blue dot-dashed),
a loop of vector-like left-handed $L$ weak doublets (blue) or of any lepton with $\U(1)_Y$ charges only (green),
provided that a production mechanism is found.
The models that violate some bounds are: 
a loop of particles with $\SU(2)_L$ charges only (black),
a loop of vector-like right-handed down quarks $D$, 
of vector-like left-handed quarks $Q$  (red dashed and magenta dotted),
and a KK graviton (red dotted).
}
\end{center}
\end{figure}

\subsection{Effective operators: spin 2}
Similar considerations hold if $\X$ has spin 2. Taking gravity as inspiration, we can couple a tensor $\X_{\mu\nu}$  to the various components $T^{(p)}_{\mu\nu}$ of the energy-momentum tensor:
\beq  \X^{\mu\nu} \sum_p \frac{T^{(p)}_{\mu\nu}}{\Lambda_p},   \eeq
where $T^{(\gamma)}_{\mu\nu} = F_{\mu\alpha} F_{\nu\beta} g^{\alpha\beta} - g_{\mu\nu}  F_{\alpha\beta} F^{\alpha\beta} /4$\, 
for a gauge boson and 
$T^{(f)}_{\mu\nu} = (\bar f \gamma_\mu \overleftrightarrow{\partial}_\nu f)/2$ for a Dirac fermion $f$. The relevant decay rates are then
{\beq
\Gamma(\X \to \gamma \gamma)  = \frac{ M^3}{ 80 \pi \Lambda_\gamma^2},\qquad
\Gamma(\X \to g g) = \frac{ M^3}{ 10 \pi \Lambda_g^2},\qquad
\Gamma(\X \to b \bar b) =  \frac{ 3M^3}{ 160 \pi \Lambda_b^2}.  
\eeq
Including the $2J+1$ factor from the 5 spin states, the signal rate is reproduced for
\beq 
\frac{M}{\Lambda_\gamma}\, \frac{M}{\Lambda_g} \approx 0.04\, \sqrt{\frac{\Gamma}{M}} \approx 0.01
\qquad{\rm or} \qquad
\frac{M}{\Lambda_\gamma}\, \frac{M}{\Lambda_b} \approx 1.2 \, \sqrt{\frac{\Gamma}{M}} \approx 0.3 \, .
\label{eq:lambdafit2}
\eeq}
Notice that a spin 2 particle with these couplings does not decay into $Z\gamma$, unlike a spin 0 particle.
In the future,
by analysing the angular distributions of the excess diphoton events, it will be possible to distinguish a spin-2 resonance from a scalar particle. 
A candidate for heavy spin-2 resonances is the graviton in warped extra-dimensional models~\cite{RS}. 
In this case all the $\Lambda_p$ coefficients would be equal: 
the resulting $\gamma\gamma$, $gg$ rates can reproduce the diphoton excess. 
However, the universality of gravity interactions implies a peak in the dilepton spectrum 
with a cross section equal to the one in two photons.
There are no indications for a peak at $\Excess$ in  Run 2 dilepton data, which imply
the 95\% confidence level bounds
$\sigma(pp\to \ell^+\ell^-)<5\fb$ (ATLAS) and $\sigma(pp\to \ell^+\ell^-)\circa{<}3\fb$ (CMS)~\cite{data}.
Only with modifications of the minimal setup one could fit the observations.

\begin{figure}[t]
\begin{center}
$$ \includegraphics[width=0.34\textwidth]{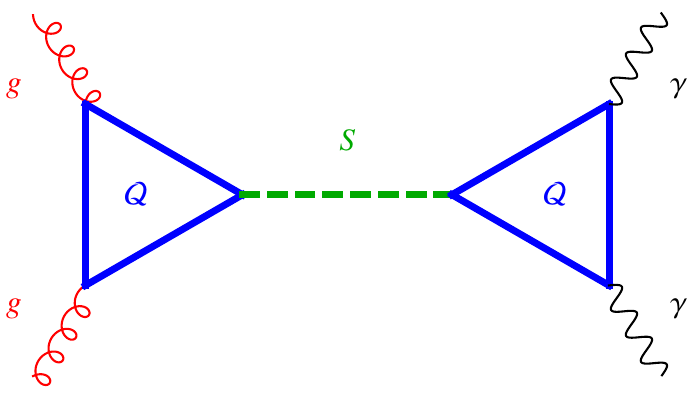}$$
\caption{\em Weakly coupled models. \label{fig:Feynloop}}
\end{center}
\end{figure}

\section{Weakly coupled models}\label{perturbative}
{Here we describe how to obtain weakly coupled (renormalizable)  models realising the scenario discussed in the previous section
via the Feynman diagram in fig.~\ref{fig:Feynloop}.
The SM is extended by adding one (or more) scalar singlets  $\X$, and
extra vector-like fermions $\Q_f$ (written in Dirac notation) 
or scalars $\tilde\Q_s$
with mass $M_i$, hypercharge $Y_i$, charge $Q_i$ and in the colour representation $r_i$,
with the couplings 
\beq \X \bar\Q_f (y_f + i \, y_{5f}\gamma_5) {\Q}_f + \X A_s \tilde\Q_s^* \tilde\Q_s
.\label{eq:YukSQ}\eeq
As before, the use of the scalar or pseudo-scalar interaction depends on the CP nature of $\X$.
This kind of structure is fairly generic in models that extend the SM sector around the weak scale.
One is easily convinced that our conclusions are not dramatically affected by allowing also matter with $\SU(2)_L$ quantum numbers.  The case in which the scalar $\X$ is part of a $\SU(2)_L$ multiplet will be dealt with later and the model building constraints imposed by the large width will be investigated in the next subsection.}

Focusing on  the CP-even couplings, we find that the fermion and scalar loops induce the following widths~\cite{nlo}:
\begin{eqnsystem}{sys:loopG}
\Gamma(\X\to gg) &=& M \frac{\alpha_3^2}{2\pi^3} \left| \sum_f I_{r_f} \sqrt{\tau_f} y_f \mathcal S(\tau_f) 
+\sum_s I_{r_s} \frac{A_s}{2M} {\cal F}(\tau_s)\right|^2\,, \\
\Gamma(\X\to \gamma\gamma) &=& M \frac{\alpha^2}{16\pi^3} \left| \sum_f d_{r_f} Q_f^2 \sqrt{\tau_f}  y_f \mathcal S(\tau_f) 
+\sum_s d_{r_s} Q_s^2 \frac{A_s}{2M} {\cal F}(\tau_s) \right|^2\,,
\end{eqnsystem}
where  $\tau_i = 4M_i^2/M^2$ and $I_r$ and $d_r$ are the index and dimension of the colour representation $r$ 
({\it e.g.} $I_3=1/2$, $I_8=3$), and
\beq
\mathcal P(\tau)   = \arctan^2(1/\sqrt{\tau-1})\ ,\qquad
\mathcal S(\tau)  = 1+ (1-\tau)\mathcal P(\tau)\,, \qquad
\mathcal F(\tau)   =\tau\mathcal P(\tau)-1 \,.
\eeq
In the limit of heavy extra particles ($\tau\to\infty$)
we have $\mathcal P(\tau) \approx 1/\tau$, $\mathcal S(\tau)\approx 2/3\tau$,
$\mathcal F(\tau) \approx 1/3\tau$
 and we obtain, for the CP-even couplings,
\begin{eqnsystem}{sys:loopGN}
\frac{\Gamma(\X\to gg)}{M} &\approx& 7.2\times 10^{-5} \left|\sum_f I_{r_f} y_f   \frac{M}{2M_f}
+\sum_s I_{r_s} \frac{A_s M}{16 M_s^2}
\right|^2 \,,\label{eq:perturbativeratesa}\\
\frac{\Gamma(\X\to \gamma\gamma)}{M} &\approx& 5.4\times 10^{-8} \left|\sum_f d_{r_f} Q_f^2 y_f \frac{M}{2M_f}
+\sum_s d_{r_s} Q_s^2 \frac{A_s M}{16 M_s^2}
\right|^2 \, ,
\label{eq:perturbativerates}
\end{eqnsystem}
where we used $\alpha_3(M/2) = 0.1$.
The effect of CP-odd interactions is obtained by replacing 
 $ y_f \mathcal S(\tau_f)  \to  y_{5f} \mathcal P(\tau_f) $ in  eq.~(\ref{sys:loopG})
 and $y_f \to 3 y_{5f}/2$ in eq.~(\ref{sys:loopGN}), and omitting the scalar contribution.  
Fig.\fig{figr} shows how various kinds of fermions contribute to the $\X\to\gamma\gamma, gg$ widths.

\medskip

\begin{figure}[t]
\begin{center}
$$ 
\includegraphics[width=\textwidth]{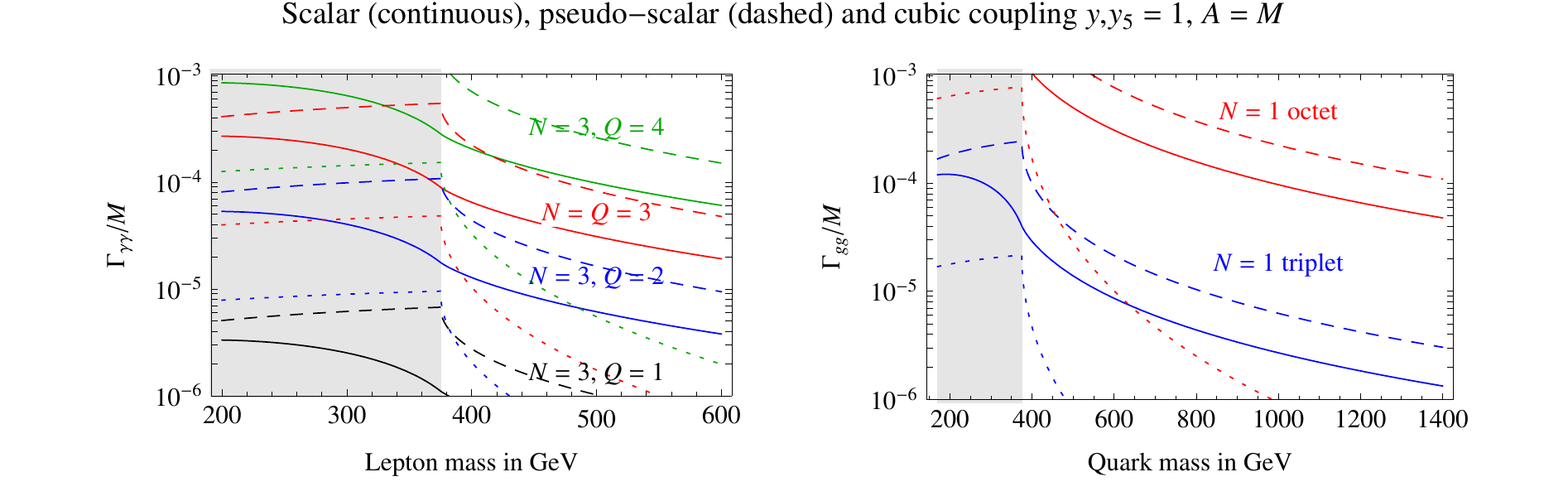}$$
\caption{\em 
$\Gamma(\X\to\gamma\gamma)/M$ and $\Gamma(\X\to gg)/M$ as generated by loops of
$N$ coloured and/or charged fermions and scalars.  In the shaded regions the fermions are lighter than $M/2=375\GeV$.
The widths grow as the square of the couplings $y_f$ and $y_{5f}$, which are taken here to be equal to 1. Continuos (dashed) curves describe the
effects produced by a scalar (pseudo-scalar) Yukawa coupling;
dotted curves describe the effect of a scalar cubic coupling $A_s=M$.
\label{fig:figr}}
\end{center}
\end{figure}

{Let us first try to explain the $\gamma\gamma$ excess rate without reproducing, at the same time, the value of the width suggested by ATLAS.  In general, allowing for other decay channels than $\gamma\gamma$ and $gg$,   the partial widths will lie in the yellow region bounded by the blue and green  bands in fig.\fig{figGG}a.
$\Gamma_{\gamma\gamma}$ is minimized when $\Gamma_{gg}$ dominates the total width, corresponding to the lower portion of the blue band and implying
  $\Gamma(\X\to \gamma\gamma)/M\approx \hbox{few}\times 10^{-6}$ and
 $\Gamma(\X\to gg)/M\approx \hbox{few}\times 10^{-3}$--$10^{-6}$.
For $y_f \sim 1$ and $M_f\sim\TeV$ such widths can be easily achieved with new matter with order one electric charges and conventional colour representations, as illustrated in fig.~\ref{fig:figr}.
Notice for instance, that a single heavy quark triplet with charge $Q$ gives $\Gamma(\X\to gg)/\Gamma(\X\to \gamma\gamma)\approx 36/Q^4$,
which ranges between $\approx 2$ and  $\approx 3000$ for $2\geq Q\ge 1/3$.
Any ratio of $\Gamma(\X\to gg)/\Gamma(\X\to \gamma\gamma)$ can be obtained by including the appropriate content of heavy leptons and quarks with different masses.
But notice that in order to reproduce $\Gamma(\X\to \gamma\gamma)/M> 10^{-6}$ using fermions of small charge, say $Q= 1/3$, a   large number
of multiplets or a large $y_f$ is needed, dangerously approaching non-perturbative dynamics, as we shall discuss below.

The masses of the required fermions can be comfortably above present bounds, depending on their representation under the SM group \cite{1404.4398}. Coloured resonances with large electric charge $Q\geq 5/3$ are strongly constrained by same-sign dilepton searches and the lower limit on their mass is of order 1 TeV, depending on $Q$ \cite{1503.05425,1401.3740}. However, as the contribution to $\Gamma(\X\to \gamma\gamma)/M$ scales like $Q^4$,
such states can easily be the dominant source of coupling compatibly with their experimental bounds. For instance, one vectorlike quark with charge $Q=5/3$ and mass $M_f=1$ TeV gives $\Gamma(\X\to \gamma\gamma)/M\sim 10^{-6} y_f^2$ and $\Gamma(\X\to gg)/M\sim 5\times 10^{-6} y_f^2$.}
On the other hand for  heavy quarks and leptons with conventional charges, the compatibility  with experimental data depends on their decay modes.
We estimate that stable charged leptons must be heavier than 0.4 TeV in order to avoid excessive Drell-Yan production~\cite{1305.0491,1504.00359}: 
thereby they cannot be lighter than $M/2$.
An exception is a vector-like lepton that fills a quasi-degenerate multiplet of $\SU(2)_L$ with a neutral component (which could be the dark matter) as lightest state.
The charged leptons decay into the neutral states emitting soft particles: such compressed spectra are only subject to LEP bounds~\cite{hep-ex/0210043} of about $100\GeV$.
Bounds on quasi-stable charged coloured particles are nontrivial to interpret. Existing searches rely on modelling of hadronization and nuclear and electromagnetic interactions within the detector, which depend a lot on the colour and charge assignments of the states. 
The search strategy and resulting constraints also depend crucially on the lifetime of the states. 
Typical bounds range from a few hundred GeV to a TeV.\footnote{Currently, the only existing searches of this kind are for stable $R$-hadrons in supersymmetric models. 
Gluinos for example are excluded up to $M_{\tilde g}> 1.2$~TeV if they decay outside the detector, while stops are excluded up to $m_{\tilde t} > 0.9$~TeV~\cite{1305.0491, 1411.6795}. If instead coloured particles are stopped within the detector and then decay, the bounds are weaker:  $m_{\tilde g}> 0.9$~TeV for gluinos, $m_{\tilde t}> 0.5$~TeV for stops and $m_{\tilde b}> 0.4$~TeV for sbottoms, with certain assumptions on the dominant decay modes and for lifetimes $1\,\mu{\rm s}<\tau<1\,{\rm ks}$~\cite{1501.05603, 1310.6584}. }

\subsection{Reproducing the total width: tree-level decays}\label{treedec}
{As shown in the previous section, it is fairly easy for weakly-coupled new physics to reproduce the values of $\Gamma(\X\to\gamma\gamma ,gg)$ required to explain the diphoton rate.
However, the total width, generated by the one-loop processes alone, is much too small to fit the value preferred by ATLAS (green band in fig.\fig{figGG}a). In this section we will explore the possibility of explaining the large width with tree-level decays.

Given a trilinear coupling $y$, the two-body    decay width is roughly $\Gamma/M\sim y^2/8\pi$;
so the relatively large total width $\Gamma/M\approx \GM$ 
can be reproduced through a tree-level decay if the relevant coupling $y$ is of order one. Reproducing
the total width forces us into different model building directions depending on whether or not $\X$ is a $\SU(2)_L$ singlet.


\subsubsection{$\X$ as a SU(2)$_L$ singlet}
The simplest  option is to have $\X$ decay to a pair $\chi\chi$ of invisible new fermions or scalars. As discussed in section~\ref{DM}, for a suitably tuned mass $2M_\chi \simeq M$, the new states would even possess the correct relic abundance to explain dark matter, and all that while remaining in the weakly  coupled  domain.  On the other hand, the constraints from the corresponding final states displayed in Table 1, place a rather strong lower bound on the width into photons $\Gamma(\X\to \gamma\gamma)/M> 2\times 10^{-4}$. By considering eq.~(\ref{eq:perturbativerates}) one finds that more extreme parameters are now needed. For instance, sticking to the case of $N_f$ quarks with  $Q=5/3$ and $M_f=1$ TeV one finds the constraint $N_f y_f\gsim 10$. 

In the absence of new light states, the large width must be accounted for by SM final states. The only renormalizable tree-level decay into SM particles
is into $hh$ and $VV$ mediated by the $S|H|^2$ operator of eq.~(\ref{eq:opsSU2}) (decays into Dark-Matter are discussed in section~\ref{DM}). Again $\Gamma/M\approx 0.06$ is reproduced for a reasonable value $\Lambda_S/M\approx 1.2$ of  the effective trilinear coupling constant, well within the perturbative domain. However, the constraints from the corresponding final states  in Table 1, imply the even stronger  bound  $\Gamma(\X\to \gamma\gamma)/M> 2\times 10^{-3}$. 
{The parameters needed in eq.~(\ref{eq:perturbativerates}) are then accordingly more extreme and less plausible. 
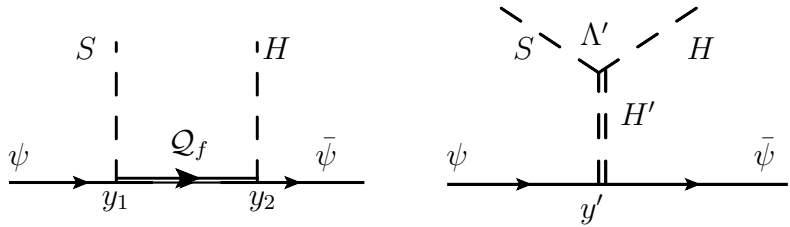
\begin{figure}
\begin{center}
\begin{picture}(155,80)(-13,0)
\put(0,0){\includegraphics[height=2cm]{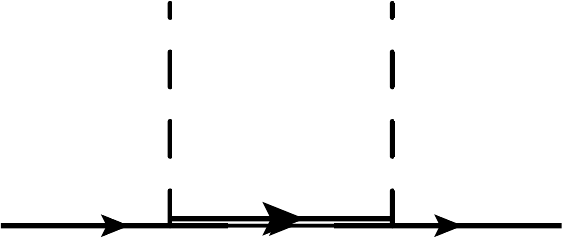} }
\put(25,50){$S$}
\put(0,10){$\psi$}
\put(60,15){$\Q_f$}
\put(35,-5){$y_1$}
\put(90,-5){$y_2$}
\put(95,50){$H$}
\put(115,10){$\bar \psi$}
\end{picture}
\hspace{5mm}
\begin{picture}(155,80)
\put(0,0){\includegraphics[width=4.5cm]{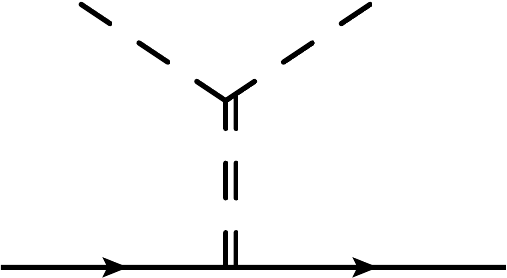} }
\put(25,50){$S$}
\put(0,10){$\psi$}
\put(50,-10){$y^\prime$}
\put(50,55){$\Lambda^\prime$}
\put(65,25){$H^\prime$}
\put(90,50){$H$}
\put(115,10){$\bar \psi$}
\end{picture}
\end{center}
\caption{\em Tree-level diagrams coupling reproducing the effective coupling between $S$ and the SM fermions $\psi$ as described in eq.~(\ref{eq:yuks}).}\label{fig:FeynDiag}
\end{figure}
For instance, sticking to  $Q=5/3$ and $M_f=1$ TeV, one now needs $N_f y_f\gsim 30$. The final remaining option is the decay to SM
fermions which is mediated by dimension 5 effective operators, see eq.~(\ref{eq:opsSU2}). As shown in fig.~\ref{fig:FeynDiag}, these can be generated either by mixing the SM fermions with heavy vectorlike counterparts or via the exchange of a heavy scalar doublet $H'$. The effective  Yukawa coupling is given in the two cases by
\be\label{eq:yuks}
y\equiv \frac{v}{\Lambda_\psi}= \left\{\begin{array}{ll}
\sfrac{y_1 y_2 v}{M_f}&{\rm (quark~mixing)}\\ 
\sfrac{y' \Lambda' v}{M_{H'}^2}  &{\rm (scalar~mixing)}
\end{array}\right.
\, .
\ee
$\Gamma/M\approx 0.06$ implies $y \approx 1$ for the decay into a quark.} We thus have two basic options:   either the new states are below a few hundred GeV, where their production and decay must be hidden by some clever model building, or, if they are at a TeV or above, 
 at least one of the trilinears $y_1,y_2, y'$ or $\Lambda'/M$ must be substantially larger than one, say $\sim 3\div 4$. This second option would push us in the strongly coupled domain. Since the required coupling to fermions is anyway substantial, naturalness suggests that the third quark family dominates 
 this final state.  Table 1 shows a constraint comparable to the case of an invisible final state: $\Gamma(\X\to \gamma\gamma)/M> 2\times 10^{-4}$.

\begin{figure}[ht!]
$$\includegraphics[width=0.45\textwidth]{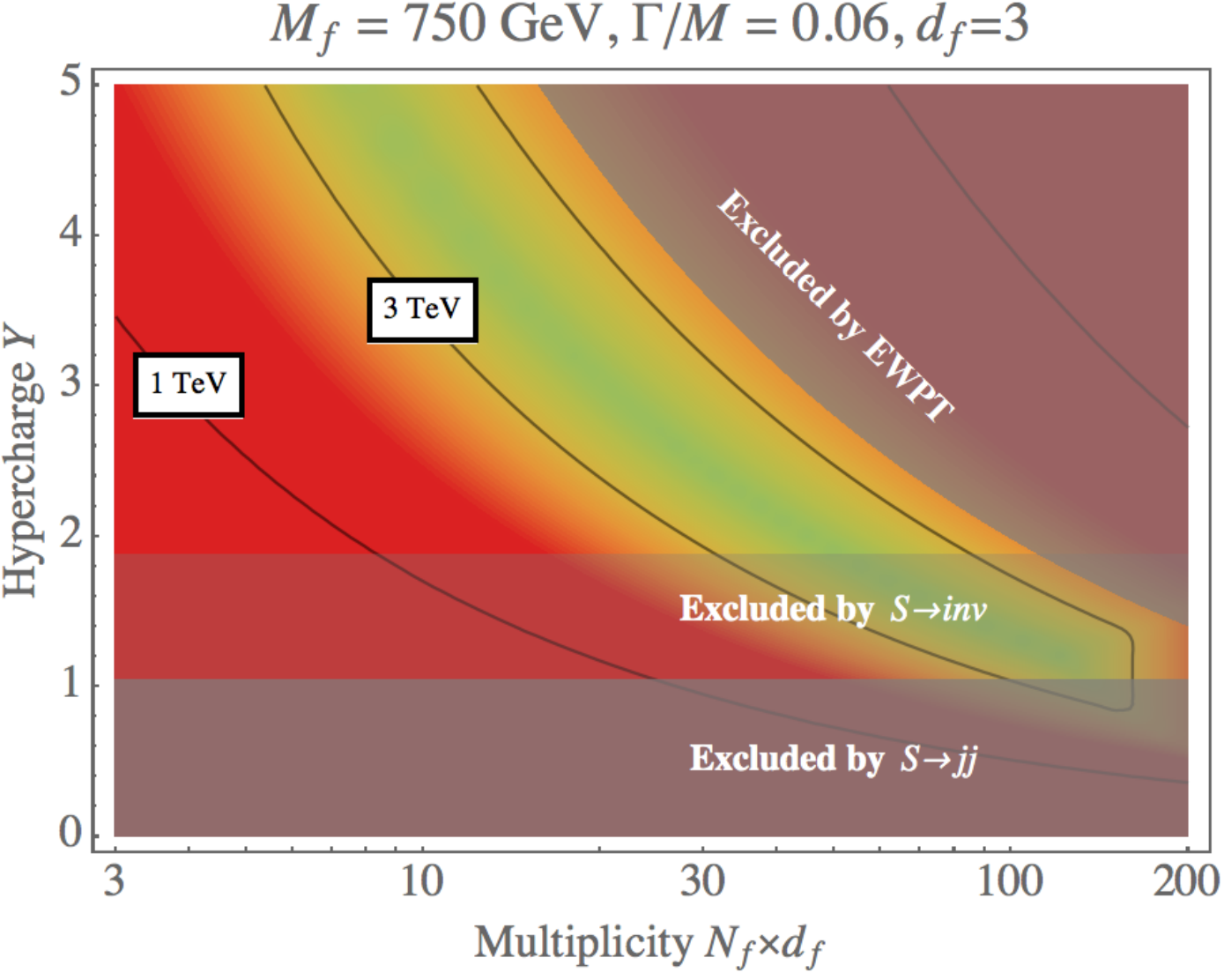}\hspace{0.5cm}
\includegraphics[width=0.45\textwidth]{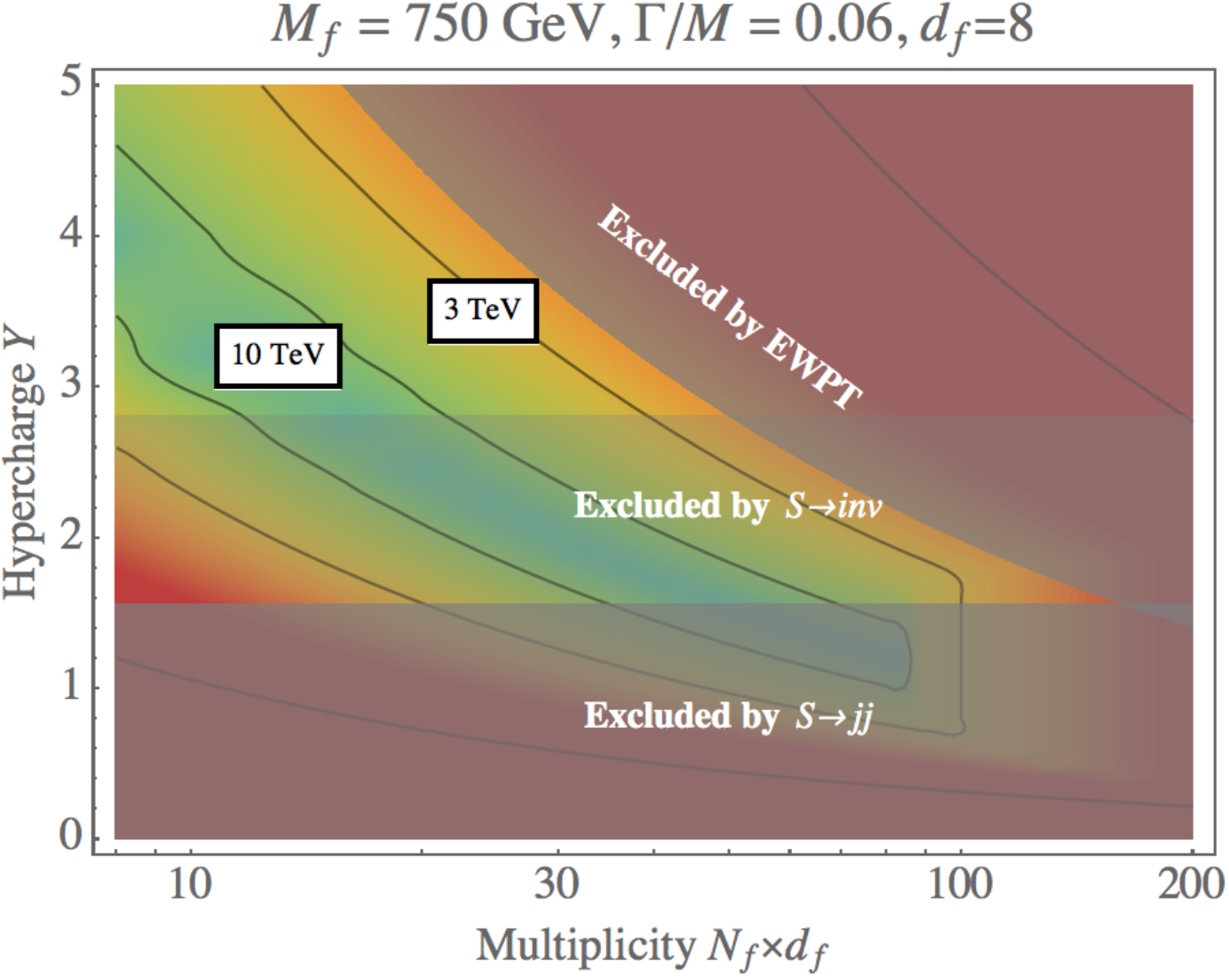}$$
$$\includegraphics[width=0.45\textwidth]{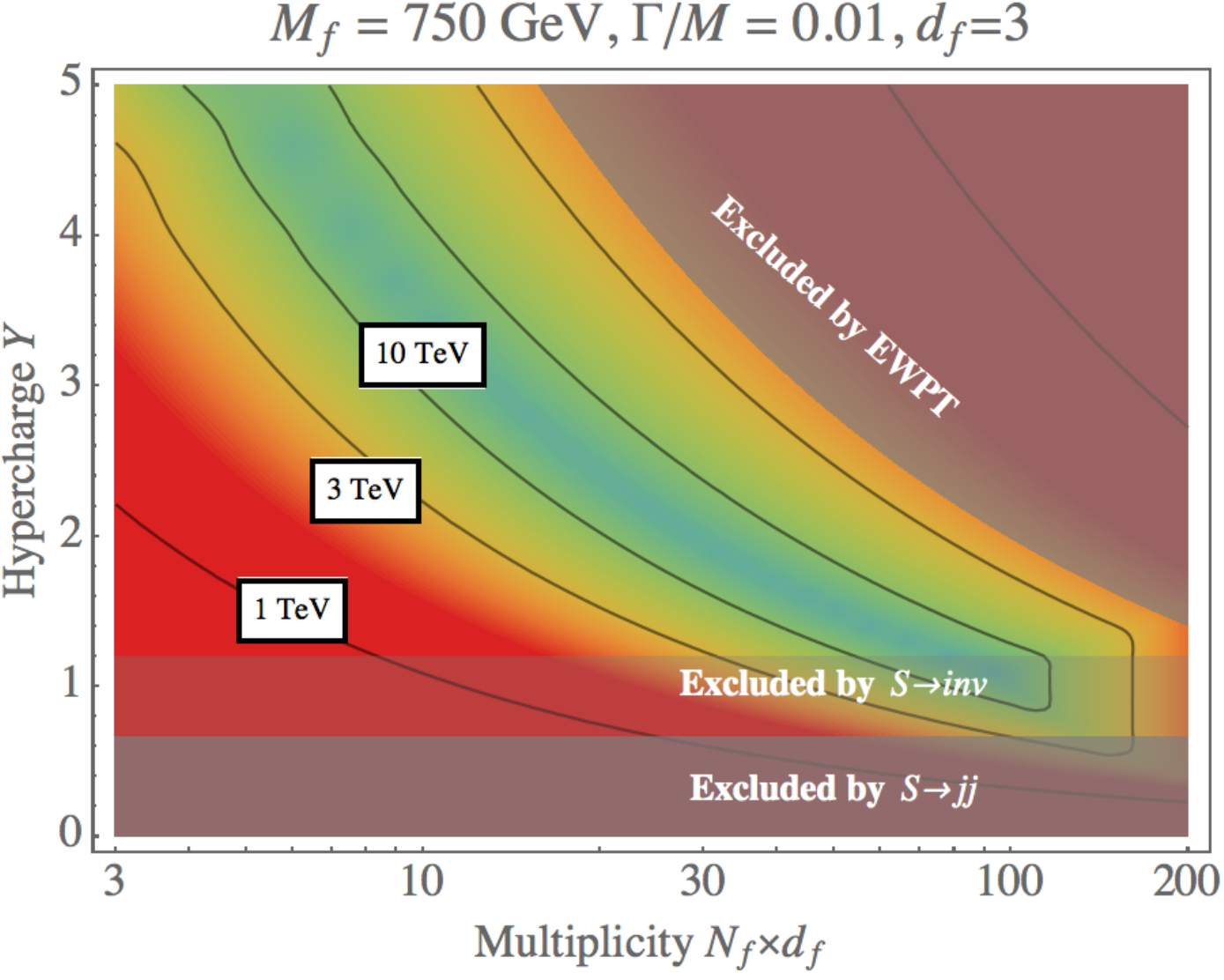}\hspace{0.5cm}
\includegraphics[width=0.45\textwidth]{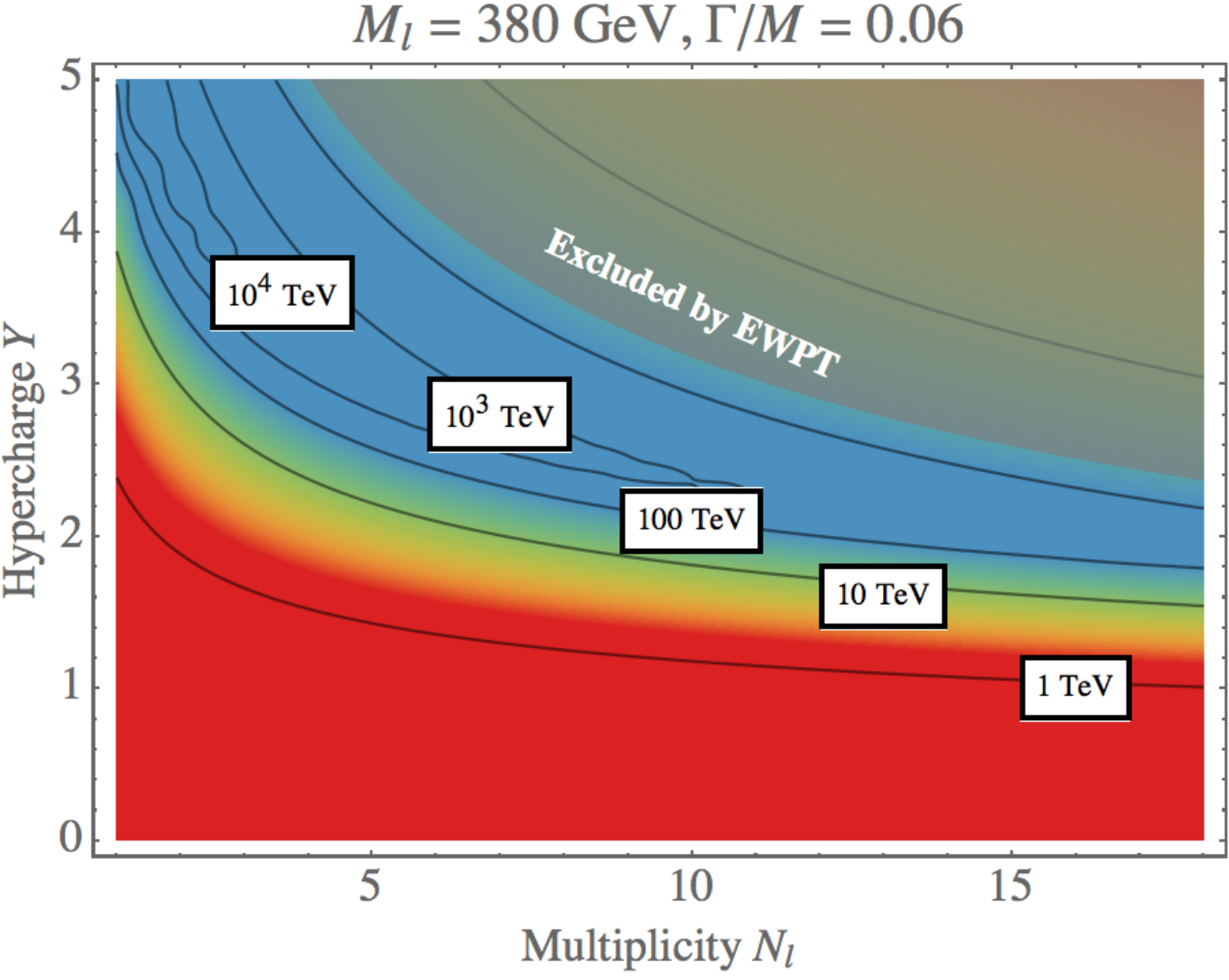}$$
\caption{\em Contour plot of the scale $\Lambda_{\rm CUT}$ where a model  with a scalar $S$ coupled to $N_f$ fermions $\Q_f$ of mass $M_f$ in the color representation $d_f$,  becomes non-perturbative. Upper plots assume $\Gamma/M=0.06$ and display the effect of fermions in the fundamental (left) and adjoint (right) color representation. Shaded bands correspond to the constraints from $pp\to\X\to \hbox{invisible}$ ($pp\to\X\to jj$) assuming the width is reproduced by invisible decays  (light) or decays into jets (dark) and to constraints from the EW {Y}-parameter (dark). The coupling that becomes non-perturbative is $g_3$ for large $N_f$, $g'$ for large $Y$, 
and the Yukawa for small $N_f Y^2$. Lower-left: smaller width $\Gamma/M=0.01$. Lower-right: $S$ couples to $N_l$ light colourless fermions and gives a partial width compatible with data if the total width is accounted for by invisible decays.
 \label{fig:RGE}}
\end{figure}

\subsubsection*{RGE running and Landau poles}

From  the above discussion, it appears that an invisible final state offers the most favorable option to remain within the weakly coupled domain. Nonetheless,  the absence of the corresponding signal in the data, requires  $\Gamma(\X\to \gamma\gamma)/M> 2\times 10^{-4}$ which in turn seems to force the couplings and quantum numbers of the underlying new states to the boundary of the weakly coupled domain. It is worth investigating the issue more quantitatively by considering the  Renormalization Group evolution  of the couplings above the weak scale (see also \cite{RGE}). We are going to focus on  specific examples, but that will be sufficient to draw general lessons.
Consider first the  
 perturbative model of section \ref{perturbative}, with $N_f$ 
fermions in the same $\SU(3)_c$ irreducible representation $f$, with hypercharge $Y$ and mass $M_f\gtrsim 750\GeV$ and universal Yukawa coupling $y_f$ to $\X$. 
The Renormalization Group Equations (RGEs) for Yukawas and gauge couplings in this model  read 
\begin{eqnarray}
16\pi^2\frac{d y_f}{dt}&=&(2 N_f d_f + 3) y_f^3 - 
6C_f  g_3^2 y_f - 6 g'^2 Y^2 y_f\,,\\
 16\pi^2\frac{d g_3}{dt}&=&\left(\frac{4N_f I_f}{3}-7\right)g_3^3\,,\\
 16\pi^2\frac{d g^\prime}{dt}&=&\left(\frac{41}{6} + \frac{4}{3} Y^2 N_fd_f\right)g^{\prime\, 3}\,,
\end{eqnarray}
where $d_f$, $C_f$ and $I_f$ are respectively the dimension, Casimir and index of the irreducible colour representation, satisfying $C_f d_f = 8 I_f$, and $t=\ln(\mu/M)$.
The  observed rate and the lower bound on $\Gamma(\X\to \gamma\gamma)/M$ give respectively two constraints
\be
y_f^2 I_f d_f N_f^2 Y^2 \simeq \frac{58}{\tau_{f} \mathcal S(\tau_f)^{2}}\sqrt{\frac{\Gamma/M}{0.06}}\,,\qquad y_f d_f N_f Y^2\gtrsim \frac{35}{\sqrt{\tau_{f} }\mathcal S(\tau_f)}\sqrt{\frac{\Gamma/M}{0.06}}\,,
\label{constraints}
\ee
for the quantum numbers and the Yukawa coupling renormalized at  the scale $M$. Here we considered a scalar $\X$ with $\tau_f = 4M_f^2/M^2$ and  $\mathcal S(\tau_{f})$ defined before (for a pseudo-scalar see below eq.~(\ref{eq:perturbativerates})). 
{Fixing $y_f$ with the first of eq.~(\ref{constraints}), the second equation as well as the demand of perturbativity  from the RG equations can be conveniently represented as constraints in the $( N_f',Y)$ plane, with $N_f'\equiv N_f d_f$ the total number of degrees of freedom.
In the upper-left panel of fig.\fig{RGE}  we present the results for the simplest case where $f$ is a colour triplet ($d_f=3$), showing that in the bulk of parameter space the UV cut-off is below a few TeV.
 Notice in particular that the second eq.~(\ref{constraints}) can  be written as a lower bound on $Y/\sqrt{C_f}$ (light-shaded region in the plots). 
Starting from this case one can 
see how things scale for a general colour representations $d_f$. By inspecting the RG equation and using eq.~(\ref{constraints}) to solve for $y_f$, 
 the constraints from the absence of Landau poles  have the form\footnote{These expressions assume that the couplings are perturbative at the scale $M$ and the bound on the Yukawa  neglects the effect of gauge couplings on the Yukawa RG: a  good approximation in the region where the relative bound dominates.}
\beq\begin{array}{rcll}
N_f' Y^2 C_f &\gtrsim& \displaystyle \frac{6\, t_{\textrm{CUT}} }{\sqrt{\tau_{f}} \mathcal S(\tau_f)}\sqrt{\frac{\Gamma/M}{0.06}} \,\,\,\, &{\rm (Yukawa)}\,,\\
N_f'C_f & \lesssim&\displaystyle\frac{24\pi}{\alpha_s\, t_{\textrm{CUT}}}+42\,\,\,\,&{\rm (strong~coupling)}\,,\\
N_f' Y^2 &\lesssim&\displaystyle \frac{3\pi}{\alpha\, t_{\textrm{CUT}}}-5\,\,\,\,&{\rm (hypercharge~coupling)}\, 
\end{array}\label{boundgprime}
\eeq
with $t_{\textrm{CUT}}=\ln(\Lambda_{\textrm{CUT}}/M)$ and $\Lambda_{\textrm{CUT}}$ the scale where the couplings become non-perturbative. Considering that $\Gamma(\X\to \gamma\gamma)/M$ places a lower bound on $Y/\sqrt{C_f}$, we see that by increasing the Casimir $C_f$ all bounds, besides the one from the Yukawa RG, are either unaffected or made stronger. The end result is that for representations with larger $C_f$ the cut-off can be extended to several TeVs,  in the region with large $Y$ and small $N_f'$, as clarified by the comparison of the upper plots of fig.\fig{RGE}, which show the cases $C_{f}=4/3$ ($d_{f}=3$) and $C_{f}=3$ ($d_{f}=8$). Notice however that, for large $d_{f}$, the second constraint in eq.~(\ref{constraints}) becomes stronger. 

Furthermore, the extra fermions affect electroweak precision observables contributing to the ${Y}$-parameter as
$N'_f  Y^2 \alpha_Y M_W^2/15\pi M_f^2$, which is experimentally constrained to be smaller than about~$10^{-3}$~\cite{STWY}
(a future $e^-e^+$ collider can significantly improve on this).
Direct constraints from  $pp\to\X\to \hbox{invisible}$ ($pp\to\X\to jj$) also provide bounds on the hypercharge $Y$, if one uses the right-hand-side of eq.~(\ref{constraints}) to solve for $y_f$ and assumes the width is reproduced by invisible decays (or decays into jets).
Finally, very large values of the hypercharge  give too large a $\Gamma_{\gamma\gamma}$, in conflict with eq.~(\ref{gagaconst}).

\medskip

The problem of these simplest models is that coloured fermions must have a mass $\gsim$ TeV, with the consequent need of sizeable Yukawa to reproduce the rates. This constraint can be relaxed by considering a model where the coupling to photons is dominantly generated by massive leptons, whose mass, compatibly with LHC observations, could be as light as $300\GeV$. From the perspective of a simple theory of EWSB this seems like a more ad hoc option, but it may be well motivated when trying to account for dark matter. 
The pertrubativity range of a model  with $N_f$ massive leptons with hypercharge $Y$
 is shown in the lower-right panel of fig.\fig{RGE}, having fixed $\Gamma(\X\to \gamma\gamma)/M=2 \times 10^{-4}$.
   Notice that in this figure only the Yukawa and hypercharge RG play a role, as the RGE for $g_3$ is not affected, and the coupling to gluons can be taken care by  quarks in the TeV range without major constraints.
For $N_f Y^2\simeq 20$ the cut-off can be pushed up to $10^{4}$ TeV or more, although 
in the bulk of the parameter space the cut off is again below 100 TeV.}

 }


\subsubsection{$\X$ in an SU(2)$_L$ doublet}

%

An interesting possibility is to consider $S$ as a neutral component of a second Higgs doublet $H'$ instead of the singlet that we have considered so far. We have in mind an inert doublet whose scalar potential respects an accidental $Z_2$ symmetry under which $H'$ is odd. The discrete symmetry is explicitly broken only by Yukawa interactions and guarantees the smallness of $\langle H' \rangle =v'=v/\tan\beta$. The decay width into $\psi=t,b$ is
\beq
\frac{\Gamma(\X\to \psi\bar \psi)}{M} \simeq \frac{3}{16\pi} y^{\prime 2}_\psi,
\eeq
where $y'_t$ and $y'_b$ are the new Yukawa couplings, which are not related to quark masses.
The required total width is obtained for $y'_t$ or $y'_b\sim 1$.
If $y'_b$ is of order one, the $H'$ contribution to the bottom mass is naturally small, as long as $\tan\beta \circa{>}y'_b/y_b \approx 50$.
Then the decay widths into massive SM vectors are safely small, as they are suppressed by $v'^2$:
\beq \frac{\Gamma(\X\to W^+W^-)}{M} \simeq \frac{2\, \Gamma(\X\to ZZ)}{M} \simeq \frac{(g_2^2+g_1^2)^2 v^{2}}{32\pi M^2 \tan^2\beta}\approx
\frac{6\times 10^{-5}}{\tan^2\beta}.\eeq
However the decay width into photons (and gluons) induced by SM fermion loops is also small, 
\beq \frac{\Gamma(\X\to \gamma\gamma)}{M} \approx \frac{y_t^{\prime 2} m_t^2}{(4\pi)^5 M^2}\label{eq:H'gamma}.
\eeq 
Indeed, while the dimension-5 effective operators in eq.\eq{ops} is electroweak gauge invariant for a scalar singlet $\X$,
a scalar doublet $H'$ must instead be contracted with the ordinary Higgs doublet, leading to a dimension-6 operator.
The result is still given by eqs.~(\ref{sys:loopG}), where the extra suppression is encoded in the fermion mass and explains the additional suppression factor $m_t^2/M^2\approx 0.05$ in eq.\eq{H'gamma}. In order to achieve the required partial width into photons, one needs 
interactions between $H'$ and extra fermions with  multiplicities and electric charges even larger than for the singlet, incurring in problems that are even more severe (w.r.t. RGE evolution and EWPT constraints) than the ones described above.

\subsection{Reproducing the total `width': multiple states}\label{many}
The difficulties in producing a large total width in weakly-coupled theories prompt us to look for alternative routes to explain the ATLAS observations.
Given that present data are not accurate enough to tell if the excess at $\Excess$ has a Breit-Wigner shape,
we can speculate that the spread observed by ATLAS is not due to a width but to two or more almost overlapping narrow resonances with comparable masses around $M\approx \Excess$ and mass differences of the order of the measured width.
The previous phenomenological analyses are easily adapted, 
by reinterpreting all $\Gamma$ as sums over the resonances, and
by ignoring the constraint on the total width $\Gamma/M$.
Then, the values of  $\Gamma (\X\to \gamma\gamma, gg)$ needed to reproduce the signal rate
can be obtained through loops of fermions or scalars, without invoking large charges or multiplicities,
and without creating any conflict with the $pp\to jj$ bound (see fig.\fig{figGG}a).

\medskip

The new model-building issue is to justify the quasi-degeneracy of the multiple states forming the observed resonance.

If the resonance is interpreted as a new Higgs doublet (or, in more generality, as a scalar weak multiplet), the near mass degeneracy is actually an automatic feature of the theory. The scalar and pseudo-scalar components of the heavy neutral Higgs are split only by electroweak effects.
For example the $Z_2$-invariant quartic interaction $\lambda [(H^\dagger H')^2+\hbox{h.c.}]/2$
gives a mass splitting $\Delta M = \lambda v^2 /M= \lambda \, (\Excess /M) \, 40$~GeV, which is of the correct size for $\lambda$ of order unity.
In the Minimal Supersymmetric SM the mass splitting is $\Delta M \simeq\frac12 M_Z^2 \sin^22\beta$, which is smaller than 6 GeV and suppressed at large $\tan\beta$.
The charged components of $H'$ can decay into $W\gamma$ and $WZ$ but cannot be singly produced through $gg$ partons.  
Production through $u\bar d$ partons is possible, if $H'$ couples to light generations.
Furthermore, scalars with electroweak interactions and $\Excess$ mass can also be pair produced, with a cross section $\sigma \sim 0.2 \fb$. Given that it is no longer necessary to have a large width, the difficulties encountered in section~\ref{treedec} disappear. Acceptable values of the decay widths in $\gamma\gamma$ and $gg$  can be reproduced by moderate couplings of $H'$ to extra matter 
with reasonable charges and masses larger than $M/2$.

If the resonances $\X$ are singlets under $\SU(2)_L$, a similar scheme can be applied. 
However, in this case having mass differences of the order of the apparent width is not an automatic feature and could require a certain degree of coincidence.
The quasi-degeneracy of multiple states could be justified by introducing extra symmetries, 
softly broken, for example,  by the masses of the fermions $\Q$. 

If the diphoton excess persists when more data is collected at the LHC, experiments will have the energy resolution needed to
distinguish a genuine large width from the multiple-state solution through a more accurate determination of the shape of the peak~\cite{1301.0328}.

\begin{figure}[t]
\centering
\includegraphics[width=0.6\textwidth]{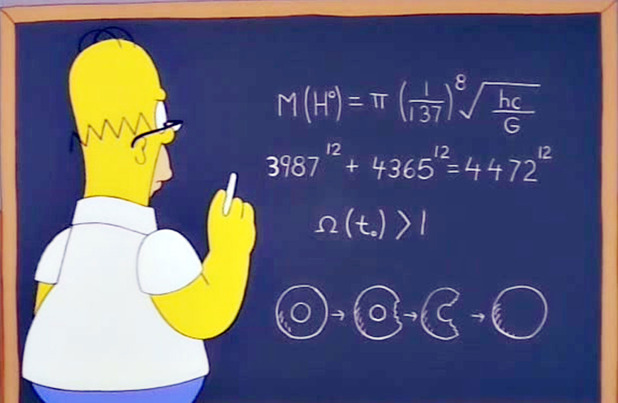}
\caption{\em 
Compilation of values of $\Gamma/M$ for bosonic neutral resonances produced by QCD interactions.
 \label{fig:bestiario}  }
\end{figure}

\section{Strongly coupled models}\label{nonpert}

As discussed in detail in the previous section,
a relatively large width $\Gamma/M\approx \GM$, combined with constraints from the rates in the various channels, would severely limit the
weakly coupled options. If one demands a weakly coupled description up to above 10 TeV, one has basically two options: in one the width is dominated by an invisible channel and light leptons at the threshold of discovery are responsible for the sizable coupling to photons (fig.~\ref{fig:RGE}, lower-right panel), in the other the width is mimicked by the presence of one or more nearby resonances.
It makes sense to investigate in more detail how well the properties of the new resonance fit scenarios with a novel strong dynamics around the weak scale (fig.\fig{bestiario} shows for the sole  purpose of illustration, with no deep meaning attached, a compilation of the $\Gamma/M$ values of the bosonic neutral resonances produced by QCD interactions).
We shall explore various incarnations of the scalar $\X$ as a composite state of the  new dynamics. 
Two broad scenarios can be imagined.
\begin{itemize}
\item $S$ is a component of an extended sector, explaining the naturalness of the electroweak scale and  producing the Higgs doublet as a composite state. 

\item $S$ belongs to a sector that is not directly responsible for electroweak symmetry breaking. 
This can be realised in explicit simple models with extra vector-like fermions, described by a QCD-like fundamental Lagrangian. 
The electroweak scale could be linked to the new strong interactions in a more subtle way as in ~\cite{strongEW} or \cite{relaxion} or because of dark matter.
\end{itemize}
A particularly motivated option is that $S$ is a light Pseudo-Nambu-Goldstone Boson (PNGB) from the spontaneous breakdown of an internal  global symmetry. Another perhaps more exotic option is that $\X$ 
is a pseudo-dilaton from the spontaneous breakdown of approximate scale invariance.  In addition one should also entertain the possibility of composite states that 
appear around the confinement scale, analogous to  charmonium in QCD.  If the new sector contains new coloured states, as most likely required
to produce $S$ from gluon fusion, coloured resonances are expected around the same scale as $S$ which could produce observable experimental signals.

\subsection{Scalars and pseudo-scalars  in  strongly-coupled models}

Strongly-coupled models at the TeV scale are mainly motivated by the hierarchy problem, which is solved
by assuming that the  Higgs  is  a composite state. 
To account for the hierarchy, the simplest picture  is that above the weak scale the new dynamics
flows to a fixed point, which can be either free, like in QCD, or interacting. The second option seems more likely to be the case in view of the need for operators with sizeable anomalous dimensions in order to account for flavour, through partial compositeness. 
Partial compositeness may be dispensed with for the light generations, but seems unavoidable to account for the sizeable top Yukawa. One remarkable consequence of partial compositeness is that the new strong dynamics must necessarily involve states  charged under all of the SM gauge group factors:
the coupling of resonances to both gluons and photons is thus an unavoidable consequence of a plausible flavour structure. It should be emphasised that minimal technicolour does not imply couplings to gluons. 

To get the simplest  and roughest idea of the dynamics from the new sector, one can imagine it to be broadly described by two parameters~\cite{Giudice:2007fh}: an overall mass scale $m_*$ and a coupling $g_*$. For instance in large $N$ gauge theories, $m_*$ coincides with the hadron mass scale and $g_*\sim 4\pi/{\sqrt N}$ with the coupling among mesons, or the topological expansion parameter. The same simple structure appears in holographic realisations of strongly-coupled theories where the two parameters are simply given by  the KK mass and coupling. In all cases $N\sim (4\pi/g_*)^2$ roughly counts the number of degrees of freedom in the new dynamics. This perhaps oversimplified picture will be the basis of our discussion.

The    lightness of the Higgs boson with respect to the strong scale $m_*\sim $ TeV is more naturally explained if
it emerges as a PNGB  from the spontaneous breakdown of a global symmetry $G \to H$. 
It is however often the case that extra  PNGB  scalars  accompany the Higgs \cite{extraPGB},
with their quantum numbers  fixed by group theory.  
For example, in  composite Higgs models based on the coset  $\SO(6)/\SO(5)$ \cite{SO6},  one obtains an extra singlet PNGB. 
Similarly to the Higgs multiplet, these extra PNGB  acquire a mass from the explicit breaking of $G$. 

Two cases for the breaking can be envisaged:
 either the breaking comes from  external sources (normally associated with the SM fields),
or  it originates from the  strong dynamics itself as we explain below.
In the first case the most generic expectation is that the PNGB mass is dominated, as for the Higgs, by quantum corrections associated with the top quark Yukawa
coupling. In the absence of tuning, we thus expect 
\begin{equation}
m_{\rm PNGB} \sim m_h \times \frac f v\sim  \frac{y_t}{4\pi}m_*\,,
\label{mass1}
\end{equation}
where $f$ is the Higgs decay constant, which is  parametrically related to the mass scale by  $m_*\simeq g_* f= 4\pi f/\sqrt{N}$. 
The absence of observable  deviations from the SM in both Higgs couplings   and  electroweak data suggests $f\gg v$. Overall $f\gsim 3 v$ seems like a fair
lower bound to meet to those constraints, while a larger separation of scales seems less plausible as $v^2/f^2 $ is a good measure of fine tuning. In view of all that,
$ 750$ GeV fits the mass of a new PNGB dominated by top loops for $v^2/f^2\sim 0.03$, a slightly worse tuning  than normally required by precision physics.

Let us consider now the second option.
One can easily think of two basic examples.
The first is given by the analogue of  the $\eta'$ in QCD like theories at large $N_c$: the anomaly, which explicitly breaks the associated $\U(1)_A$,
is a subleading effect in the $1/N_c$ expansion. Consequently the mass of the PNGB is \cite{Witten:1979vv}
\begin{equation}
m_{\rm PNGB} \sim \sqrt{\frac{N_f}{N_c}} m_*\,,
\label{massone}
\end{equation}
where $N_f$ is the number of flavours. 
A second example arises in a 5-dimensional model and consists of a Wilson line $W=\oint A_5$ associated with a 5-dimensional $\U(1)$ gauge field, $A_M$, $M=(\mu,5)$, whose mass vanishes at tree level and generally arises at the one-loop level 
from the Aharonov-Bohm effect. The resulting mass, according to our identification of parameters, would scale like
\be
m_{\rm PNGB} \sim \frac{g_*}{4\pi} m_*\,.
\label{mass2}
\ee
This 
could easily appear in   holographic realisations of the composite Higgs. In that case the SM gauge fields $A^a_M$  also propagate in the fifth-dimension. A 5D Chern-Simons term could then couple the $\U(1)$ gauge field $A_M$ to  the SM bulk gauge fields
\be
\epsilon^{MNRST} A_M F^a_{NR} F^a_{ST}\,.
\ee
After compactification this would give rise to a pseudoscalar coupling between the Wilson line scalar $W$ and the 
4D SM gauge fields, photons and gluons in particular. Notice, as a structural remark, that  by interpreting $g_* \sim 4\pi/\sqrt{N_c}$, 
eq.~(\ref{massone}) and eq.~(\ref{mass2})   coincide for $N_f \sim {\cal O}(1)$,  which is nice and reassuring.
One should however remark that when $N_f\sim N_c$ there is no parametric suppression for the $\eta'$ mass. 
Now, if the theory underlying electroweak breaking is a gauge theory in the conformal window,  one would expect $N_f\sim N_c$. If  that is the case, one would then have to imagine some more clever mechanism to explain the mild separation of scales between the PNGB and the heavier resonances. It is perhaps an interesting question of model building how to implement the analogue of the $\eta'$ in large $N$ theories sitting near a strongly coupled fixed point above the hadron mass scale.

\medskip

In any case, in composite Higgs models we have at least two options to produce PNGB scalars whose masses are in the range $m_h\ll m_{\rm PNGB}\ll m_*$, and given by eqs.~(\ref{mass1})--(\ref{mass2}). It is amusing to remark  that there exist several independent indications  that $g_*$ should be smaller than its  maximum allowed value
$\sim 4\pi$. First, the need for a rich operator structure to accommodate flavour suggests the number of degrees of freedom should be large ($N\gg 1$). Also the observed value of the Higgs mass is naturally obtained in composite Higgs models
for  $g_* \sim 2-4$ \cite{DeSimone:2012fs}.
Finally, the slight excess in di-bosons at around $2$ TeV in LHC Run 1 can be explained as the production of vector resonances again provided $g_* \sim 3$ (one must however be aware that no corresponding excess is observed at run2, even though the integrated luminosity is not yet sufficient to deem the run1 excess  a fluctuation).
 All these facts set the stage for the existence of a scalar  whose mass, if controlled by 
eqs.~(\ref{mass1})--(\ref{mass2}), is naturally below the  trove of heavier resonances.



\medskip

Finally, and  partly mentioned above, 
an important   feature of composite Higgs models
 is the idea of partial compositeness. 
This assumes that SM fermions $f^i_{L,R}$ couple linearly to operators of the strong sector, according to
\be
\Lag_{\rm mix}=\epsilon_L^i \bar f^i_L \Psi_R^i+\epsilon_R^i \bar f^i_R \Psi_L^i\equiv \epsilon_L^i{\cal O}_L^i+ \epsilon_R^i{\cal O}_R^i\,.
\label{mix}
\ee
$\Psi_{L,R}^i$ are composite operators of dimension $\frac{3}{2}<d_{L,R}^i\leq\frac{5}{2}$, while $\epsilon^i$  parametrize their mixing with the SM fermions.
As already mentioned, a first important implication of such structure is that the strong sector must contain states that are charged under all the SM gauge interactions.
Secondly, the above mixing leads  to  a  SM Yukawa  structure given by
\begin{equation}
y^{ij}_{\rm SM} \sim \epsilon^i_L \epsilon^j_R g_*\,.
\end{equation}
This also determines the coupling the SM fermions with any state of the strong sector, our light scalar in particular.
For example in the  $\SO(6)/\SO(5)$ model~\cite{SO6}, the singlet PNGB (the $\eta$ state)
couples to  fermions as 
\begin{equation}
g_{\eta f f} \sim \frac {m_f} f  \cot \theta\,,
\end{equation}
where the angle $\theta$ depends on  the embedding of the SM singlets in the $6$ of $\SO(6)$.

In what follows we shall consider in turn the two options where $\X$ is either a scalar or a pseudo-scalar. The absence of new physics in CP violation, in particular electric dipole moments, makes it rather plausible that the strong sector  respects CP to a good degree. Because of this, our classification in terms of scalar and pseudo-scalar seems well motivated.

\subsubsection{Scalar resonances}

Consider first the case of a scalar resonance. Notice that if $\X$ is a PNGB of an internal symmetry, that very same symmetry will protect both its mass ($m_S\ll m_*$) and  its couplings to gluons and photons. Indeed, as $\X$ is colour and charge neutral, the spontaneously broken global symmetry generator associated with $\X$
must commute with $\SU(3)_c\otimes \U(1)_{\rm em}$. Therefore 
the gauging of colour and electric charge does not disrupt the Goldstone nature of $\X$. Under these circumstances the coupling to gluons and photons will feature an additional suppression, given respectively  by $y_t^2/g_*^2$ and $g_*^2/16\pi^2$ for the cases discussed in eq.~(\ref{mass1}) and eq.~(\ref{mass2}). The same suppression for the case of the PNG Higgs was discussed in ref. \cite{Giudice:2007fh}. On the other hand, if $\X$ is the dilaton associated with scale invariance (a space-time symmetry) the above conclusion does not apply. We shall devote a specific subsection to the case of the dilaton, just in view of its peculiarity and popularity.
In this section we shall instead consider 
the situation where $\X$ is an ordinary scalar resonance which happens to be accidentally   lighter than the others.
The effective Lagrangian of eq.~(\ref{eq:opsSU2}) is conveniently written as
 \bea
\Lag_{S}&=&
-\frac{1}{16\pi^{2}}\frac{S}{f}\left[
b_G\, g_{3}^{2}G_{\mu\nu}{G}^{\mu\nu}
+b_W\,  g_2^2W_{\mu\nu}{W}^{\mu\nu}
+b_B\,  g_1^{2}B_{\mu\nu}{B}^{\mu\nu}\right]+\nonumber\\
&&+c_H \frac{S}{f}|D_\mu H|^2+c_S m^2_h  \frac{S}{f}|H|^2
+c_f y_f\frac{S}{f}H\bar f_L f_R+\hbox{h.c.}\,.
\label{LSanomaly}
\eea
According to standard power counting \cite{Giudice:2007fh}, the coefficients  $ c_H$, $c_S$ and $c_f$ are expected to be $\sim 1$, while  the $b$'s are expected to be  $\sim (4\pi/g_*)^2\sim N$, where   $N$ roughly counts the number of microscopic degrees of freedom. It is interesting  to match the above effective lagrangian to that obtained by integrating out $N_F$ heavy fermion multiplets with mass $M_Q$, transforming in a representation $({d_3},{d_2})_{Y}$ under $\SU(3)_c\otimes \SU(2)\otimes \U(1)_Y$
and coupled to $\X$ with Yukawa $y$  according to  eq.~(\ref{eq:YukSQ}). One would find
\be
\frac{b_G}{f} \equiv \frac{2 }{3}  N_F d_2I_3 \frac{y}{M_Q}\qquad \quad\frac{b_W}{f} \equiv\frac{2}{3} N_F d_3 I_2  \frac{y}{M_Q} \qquad\quad \frac{b_Y}{f} \equiv\frac{2 }{3}   N_F d_3d_2 Y^2\frac{y}{M_Q}
\label{match}
\ee
where $(d_2,I_2)$ and $(d_3,I_3)$ are  dimension and index of the representation under respectively $\SU(2)_L$ and $\SU(3)_c$. This weak coupling result nicely matches  the strong coupling power counting 
by identifying $f\equiv m_*/g_*$ with  $M_Q/y $ and $(4\pi/g_*)^2$ with $N_F$. In other words, the computations of  section 3 carry over as rough estimates to the strongly coupled scenario. But rough estimates are more than suitable at the level of our analysis.
%

As long as the $b$'s are not too large, {\it i.e.} $b_i < (4\pi/\alpha_i)$, the decay of $\X$ will be dominated by 
 either $c_t$  or by $c_H$. Assuming dominance of each term in turn, we find 
\begin{equation}\label{const1}
\left\{\begin{array}{ll} \displaystyle
c_t\approx  3.5 \left(\frac{f}{M}\right)\sqrt{\frac{\Gamma}{45\GeV}}&\textrm{for}\,\,\Gamma\approx\Gamma(S\to\bar tt)\\ 
\displaystyle
 c_H\approx 2.5 \left(\frac{f}{M}\right)\sqrt{\frac{\Gamma}{45\GeV}}& \textrm{for}\,\,\Gamma\approx\Gamma(S\to W^+W^-,ZZ,hh)\end{array}\right.\,.
\end{equation}
From table \ref{tabounds}, we can read that in the former case the $\gamma\gamma$ rate is reproduced for
$\Gamma_{\gamma\gamma}/M\gtrsim2\times10^{-4}$. In the latter case, on the other hand, the constraints are dominated by the $ZZ$ final state which contributes only to 25\% of the width and imposes $\Gamma_{\gamma\gamma}/M\gtrsim2.5\times10^{-3}$: a result that  is already on the boundary of conflict with the upper limit on production from $\gamma\gamma$, eq.~(\ref{gagaconst}).
These can be written using the notation of eq.~(\ref{LSanomaly}) as
\begin{equation}\label{const4}
b_B+b_W\gtrsim \left\{\begin{array}{ll} 
\displaystyle 80\left({f}/{M}\right)\sqrt{{\Gamma}/{45\GeV}}&\textrm{for}\,\,\Gamma\approx\Gamma(S\to\bar tt)\\ 
\displaystyle  285\left({f}/{M}\right)\sqrt{{\Gamma}/{45\GeV}}& \textrm{for}\,\,\Gamma\approx\Gamma(S\to W^+W^-,ZZ,hh)\end{array}\right.\,.
\end{equation}
Finally, we can write the requirement on $\sigma(pp\to \X\to \gamma\gamma)$, 
eq.~(\ref{eq:GM}), as a relation between the coefficients
\begin{equation}\label{const3}
(b_B+b_W)b_G\approx 230\left(\frac{f}{M}\right)^2\sqrt{\frac{\Gamma}{45\GeV}}\,.
\end{equation}
which leads to
\begin{equation}\label{const2}
b_G\lesssim \left\{\begin{array}{ll} 
\displaystyle 2.8 \left({f}/{M}\right)&\textrm{for}\,\,\Gamma\approx\Gamma(S\to\bar tt)\\ 
\displaystyle 0.8\left({f}/{M}\right)& \textrm{for}\,\,\Gamma\approx\Gamma(S\to W^+W^-,ZZ,hh)\end{array}\right.\,.
\end{equation}
We can now comment on the two options. Consider first the case where $\Gamma(S\to VV,hh)$ dominates. On one hand, the correct width is reproduced for a rather reasonable value of  $c_H$ (notice indeed that $c_H= 2$ is the prediction when $S$  is interpreted as the radial mode of the $\SO(5)/\SO(4)$ $\sigma$-model). On the other hand, the required values of both $b_G$ and $(b_B+b_W)$ appear rather extreme and difficult to reconcile with one another, beside being in tension with the upper bound  eq.~(\ref{gagaconst}).

In the case where $c_t$ dominates the width, the situation is partially reversed. For $f\approx M$, corresponding to a reasonable $v^2/f^2\sim 0.1$, we find $c_t\approx 3.5$, which is somewhat on the large side, with respect to what suggested by model building practice. 
However in this case the values of $b_G$ and $(b_B+b_W)$ are more consistent. For instance, making the rough estimate $b_G= 2N_F d_2 I_3/3$ and so on (in the obvious way)  in eq.~(\ref{match}), we would obtain the needed coefficients by considering $N_F=4$ fermions in the $({3,2})_2$ of the SM group, which is quite reasonable. Moreover the needed value of $c_t$  would become more reasonable 
for a smaller width: for instance $\Gamma =10 \GeV$ would imply $c_5\simeq 1.5$. Needless to say, the presence of extra light neutral states
leading to a dominant invisible width would also help making the numbers more reasonable.
%

\medskip

As a final remark notice that, in view of the rather large value of $b_W+b_B$, one generically expects  a similar enhancement factor in the coupling of the Higgs to photons. Such enhancement would partially compensate for the suppression $(y_t^2/g_*^2)(v^2/f^2)$ due to the Goldstone nature of the Higgs, though it may not necessarily require extra tuning in view of the present bounds on deviation from the SM in $h \gamma\gamma$.

\subsubsection{Pseudo-scalar  PNGB}
If the putative 750 GeV resonance is a pseudoscalar  PNGB,  then its coupling to photons and gluons 
can be generated through the anomalous breaking of the corresponding symmetry without featuring extra suppressions. That is the situation one encounters with the  
  $\eta'$ in QCD. Similarly, in holographic models the coupling could come from the 5-dimensional Chern-Simons term. In the low energy effective lagrangian the coupling to photons and gluons would arise from a Wess-Zumino term. The simplest option is that $S$ is associated with a $\U(1)_A$ factor in the global symmetry group  \footnote{It is interesting to contemplate the case where  both $S$ and the Higgs live in the coset $G/H$ of a simple group $G$. 
  Since $G_\textrm{SM}\subset H$, the simplest option    allowing for a Wess-Zumino term seems $\SU(11)/\SO(11)$.  }. The leading terms in the effective lagrangian will be
\bea
\Lag_{P}=
-\frac{1}{16\pi^{2}}\frac{S}{f}\left[
c_G\, g_{3}^{2}G_{\mu\nu}\tilde{G}^{\mu\nu}
+c_W\,  g_2^2W_{\mu\nu}\tilde{W}^{\mu\nu}
+c_B\,  g_1^2B_{\mu\nu}\tilde{B}^{\mu\nu}\right]
+i\tilde c_fy_f\frac{S}{f}H\bar f_L f_R+\hbox{h.c.}\, .
\label{LPanomaly}
\eea
As for the scalar, we expect $\tilde{c}_f\sim 1$ while the anomaly coefficients $c_{G,W,B}$ can be parametrically large in  a theory with a large number of degrees of freedom. Indeed when matching to a weakly coupled case, the coupling $c$'s to vectors satisfy an analogue of eq.~(\ref{match}) with $2/3\to 1$. 
All the results from the previous subsection apply, with the obvious substitution $b_{G,W,B}\to c_{G,W,B}$ and with the interesting fact that now $c_H=0$ due to the CP properties of $S$. Therefore, in the absence of additional light degrees of freedom, $\X$ is expected to decay into $t\bar t$, with the same conclusions of the previous section as concerns the plausibility of the needed parameters.
%

As we already mentioned, for a CP-odd $S$, Bose symmetry and $\SU(2)_L$ forbid its two body decay into  the Higgs doublet states ($hh$, and $W_LW_L/Z_LZ_L$ according to the equivalence theorem).  Further assuming that the strong sector respects an  SO(4)  global symmetry  under which the Higgs transform as a $4$ (a custodial symmetry to  avoid large effects on the $T$ parameter),
the first allowed term coupling  $S$ to  the Higgs   is
\be
\frac{S}{f^5}\epsilon^{\mu\nu\rho\sigma}
\epsilon^{IJKL}
D_\mu h_I
D_\nu h_J
D_\rho h_K
D_\sigma h_L\, ,
\ee
where $I,J,...$ label the indices under the SO(4).
This coupling  leads to the  strong dynamics decay $S\to W^+W^-Zh$, which seems unfortunately  too  small to be seen in the near future.

\subsubsection{The special case of a dilaton}
\label{sec:dilaton}

A special instance of a light composite scalar  could be provided by the dilaton. 
The case of a light dilaton seems less
theoretically robust than the case of a light scalar from an internal symmetry. For instance in large $N$ gauge theories one could consider trying to 
extend the Veneziano-Witten argument for the axial anomaly  to the anomaly of the scale current. 
As the latter has a gluon contribution, its effects are not suppressed in the $1/N$ expansion, and therefore there is no parametric reason to expect a light scalar. The difficulties in obtaining a naturally light dilaton have been outlined in~\cite{Sundrum:2003yt}. Nevertheless, as it was shown in~\cite{Coradeschi:2013gda}, under  rather special conditions it is  possible to have  in  the spectrum   a naturally light pseudo-dilaton with mass $m_D\ll m_*$. 
The minimal Goldberger-Wise stabilisation mechanism~\cite{Goldberger} within the Randall-Sundrum (RS) model~\cite{RS} is an instance of that. 
This opens the possibility to consider the  dilaton  as the 750 GeV resonance. 
The interactions of the dilaton to the SM fields are rather constrained, depending
on whether  the SM fields are composite or not.
In what follows we will describe  the effective Lagrangian for the dilaton relevant for describing its production and decay.

Let us consider first the conceptually  simplest, though least plausible, situation where all of the SM descends from the CFT and thus respects conformal invariance up to the small effects of order ${\cal O}(\epsilon_D)$ that parametrize the explicit breakdown of scale invariance. The original RS scenario is a holographic realization of that. 
The Lagrangian is easily written by embedding the dilaton $\Omega\equiv e^{\X/f_D}$ inside the metric
 $g_{\mu\nu}=\Omega^2\eta_{\mu\nu}$, and then writing the most general diffeomorphism invariant action. 
 At the two derivative level, focussing on the dilaton+Higgs+top sectors we have
 \bea
\Lag_\Omega&=& -\frac{1}{2}f_D^2m_D^2(\Omega-1)^2 + \sqrt{-g}\bigg[ \frac{f_D^2}{12} R(g)+|D_\mu \hat H|^2+\xi R(g)|\hat H|^2 +
\label{eq:Omegazero}
\\
&&+ i\bar{\hat \Psi}\!\!\not\!\! D\hat \Psi -(y_t\bar{\hat \Psi}_L\hat H \hat \Psi_R+\,{\rm h.c.})-V(|\hat H|^2)\bigg] \nonumber \,,
 \eea
 where the first term represents the leading explicit breaking of scale invariance: $m_D^2\equiv \epsilon_D m_*^2\ll m_*^2$. We have neglected all corrections of order $\epsilon_D$
 in the couplings to Higgs and top. 
 The parameter $\xi$ is allowed by conformal invariance, but breaks any shift symmetry protecting the Higgs potential. 
 Therefore in models of PNGB Higgs   $\xi$ is expected to arise at the loop level  (dominated by top loops)
and can be neglected.
 Now, by rescaling the fields $\hat H\Omega \equiv H$, $\hat \Psi \Omega ^{3/2}\equiv\Psi$,  
 eq.~(\ref{eq:Omegazero}) can be written as
\bea
\Lag _\Omega&=&\frac{f_D^2}{2}\left[(\partial_\mu \Omega)^2-m_D^2(\Omega-1)^2\right]+ \kappa \Omega^{-1}\Box \Omega |H|^2+ |D_\mu H|^2+\nonumber\\
&&+i\bar \Psi \!\!\not\!\! D\Psi -(y_t\bar{\Psi}_LH \Psi_R+\,{\rm h.c.})-\Omega^4V(\left|{H}/{\Omega}\right|^2)\,,
\label{Omega}
 \eea
 where $\kappa =1-6\xi$. Notice that, aside from the terms in the Higgs potential (quantitatively negligible for our purposes), the couplings of the dilaton are all controlled by $\kappa$, which 
 in the particular  case of a PNGB Higgs is expected to be $ \simeq 1$. 
 Now, it is easy to check that by the further field redefinition  $H\to H/(1-\kappa\X /f_D)$, and by neglecting terms that are at least quadratic in $\X$, we obtain the effective Lagrangian for the leading  1-dilaton vertices
 \be
\Lag_1= \frac{ \X}{f_D}\left [ 2\kappa |D_\mu H|^2-\kappa(y_t\bar{\Psi}_LH \Psi_R+\,{\rm h.c.})\right]\,.
 \label{final1}
 \ee
 Here we have neglected the effects of the Higgs potential: one can be easily convinced that the potential gives effects of relative size $m_h^2/m_D^2$ as compared to the first term above.

\medskip

Let us  now consider possible departures from  the above scenario.
For example, let us  consider the case in which 
 the top  and the SM gauge sector do not arise from the CFT, but are external
fields coupled to it.  A plausible situation in this case is to assume that  the top Yukawa arises from the mechanism of partial compositeness,
where  the mixings $\epsilon^t_{L,R}$ to the strongly-coupled CFT 
depends on the dimension of the operators ${\cal O}_L^t$ and ${\cal O}_R^t$,  defined in  eq.~(\ref{mix}).
In this case   
the $\Omega$-dependence in 
eq.~(\ref{Omega}) is modified  to
\be
(y_t\bar{\Psi}_LH \Psi_R+\,{\rm h.c.})\quad\rightarrow\quad (y_t\bar{\Psi}_LH \Psi_R+\,{\rm h.c.})\Omega^{-8+d_L +d_R}\,,
\ee
where  $3\leq d_{L,R}\leq 4$ 
are the dimensions of the operators ${\cal O}_{L,R}^t$.
The lower-limit on $d_{L,R}$ arises  from the unitarity  bound, while the upper-bound is to assure that
the coupling of the top to the CFT is at least marginal.
The linear effective couplings of the dilaton are now given by 
\be
\Lag_1= \frac{\X}{f_D}\left [ 2\kappa  |D_\mu H|^2-(\kappa-\Delta_t)(y_t\bar{\Psi}_LH \Psi_R+\,{\rm h.c.})\right]\,,
\label{final2}
\ee
 where $\Delta_t\equiv 8-d_L -d_R$, satisfying $ 2\geq\Delta_t\geq 0$. The above equation describes the most general plausible scenario for the leading coupling of the dilaton to the top and Higgs sector.
 
\bigskip 
 
 Let us consider finally the couplings of the dilaton to the SM gauge fields. At the leading order they  will have the general form 
 given in eq.~(\ref{LSanomaly}), with $f_D=f$.
 Now, the two scenarios considered before must be distinguished, as they give different values for the $b_{G,W,B}$ coefficients.
In the case  in which the SM gauge fields are part of the CFT, and their interactions respect conformal invariance, 
the couplings to the dilaton are dictated by the trace anomaly equation, and the $b_{G,W,B}$ of eq.~(\ref{LSanomaly}) coincide with the beta-function coefficients in the SM: $(b_B,b_W,b_G)=-\frac{1}{2}(-41/6,19/6,7)$, see~\cite{Giudice:2000av}. In the other case, in which the SM gauge fields are  external to the CFT
that  contains matter charged under the SM gauge group,  the couplings to the dilaton come from a threshold correction upon integrating out the massive resonances whose masses are effectively $m_*\Omega$. In that case  the $b_{G,W,B}$ coincide with the contributions of CFT matter to the SM beta functions
  and are model dependent \footnote{One may worry about corrections to the coupling of $\X$ to photons and gluons induced by top and Higgs loops through the coupling in eq.~(\ref{final2}).
 One is easily convinced that top loops give an effect that is suppressed by $m_t^2/m_D^2$ as compared to the generic contributions  from the beta functions, and thus are negligible.
 On the other hand, Higgs loops from the first  term in eq.~(\ref{final2}) could affect the coupling to photons without power suppression, something  that  deserves a further study.}.   The only constraints come  from unitarity that demands $b_{G,W,B}>0$.

\medskip

For $\kappa\sim 1$, as in the case of a PNGB Higgs, the  dilaton coupling to the Higgs in eq.~(\ref{final2}) 
 dominates the decay width and fixes $f_D/\kappa\sim 600$ GeV. In this case, the rates dictate the same results of eqs.~(\ref{const3}, \,\ref{const2}) (up to a $O(1)$ normalization factor): $b_G\lsim 0.6$ and  $b_W+b_B\gsim 230 $. 
 On the other hand, for $\kappa\sim 0$ and a mechanism of partial compositeness giving a large $\Delta_t$,  we can use
   the dilaton decay into top quarks to provide its large width for $f_D=400(\Delta_t/2)$~GeV. Then $b_G\sim 1.5$ and only $b_B+b_W\sim 45$ are required to obtain the correct width into photons and gluons. In the latter case it is however impossible to interpret the Higgs as a light PNGB.

\subsection{Vector-like confinement}

We now consider QCD-like strongly coupled sectors that do not break the electroweak symmetry  \cite{sundrum}.
These models were already considered in the past as ways of dynamically inducing the electroweak scale~\cite{strongEW}, 
and as ways of obtaining composite dark matter~\cite{strongDM}.  Explicit fundamental models can be easily written down.
In the minimal scenario one adds to the SM (with its elementary Higgs doublet) 
new vector-like fermions $\Q$ charged under the SM and under  a non-abelian gauge group, 
such as $\SU(\NTC)$, with a gauge coupling that becomes strong at a scale $\Lambda$.

Upon confinement such  theories give rise to composite particles of various spins that couple to SM vectors and
are compatible with present data from flavor, precision tests and direct searches even for a dynamical scale $\Lambda\circa{<} 1\TeV$ 
(unlike technicolour models where  strong dynamics breaks $\SU(2)_L$). When the fermions are lighter than the confinement scale, 
chiral symmetry breaking at the scale $f$ produces light scalar ``techni-pions'' with quantum numbers 
determined by the constituent fermions. In particular one obtains a number of singlets equal to the number 
of irreducible fermion representations.  The decay of the techni-pion singlets into SM gauge bosons is determined 
by the chiral anomalies as for $\pi_0\to \gamma\gamma$ in the SM. 
In addition other scalars such as a dilaton or generic composite states will appear in the spectrum (c.f. discussion in Sec.~\ref{sec:dilaton}).

\begin{table}
\begin{center}
\begin{tabular}{c||c|c|c|c|c|c||c|c|c|c|c|c}
& $\frac {c_B^\eta}{\NTC}$\,&  $\frac{c_W^\eta}{\NTC}$\,&$\frac{c_G^\eta}{\NTC}$\,  & $\frac{\Gamma^\eta_{\gamma Z}}{\Gamma^\eta_{\gamma\gamma}}$ & $\frac{\Gamma^\eta_{Z Z}}{\Gamma^\eta_{\gamma\gamma}}$ & $\frac{\Gamma^\eta_{GG}}{\Gamma^\eta_{\gamma\gamma}}$ & $\frac{c_B^{\eta'}}{\NTC}$\,& $\frac{c_W^{\eta'}}{\NTC}$\,& $\frac{c_G^{\eta'}}{\NTC} $\,  & $\frac{\Gamma^{\eta'}_{\gamma Z}}{\Gamma^{\eta'}_{\gamma\gamma}} $ & $\frac{\Gamma^{\eta'}_{Z Z}}{\Gamma^{\eta'}_{\gamma\gamma}}$ & $\frac{\Gamma^{\eta'}_{GG}}{\Gamma^{\eta'}_{\gamma\gamma}}$ \,\\  \hline
$ D+L$		&  $\frac 1 {6}\sqrt{\frac 5 3}$ &  $\frac 1  2\sqrt{\frac 3 5}$  & -$\frac 1{\sqrt{15}}$ & 1.8 & 4.7 & 240 & $\frac 1 3  \sqrt{\frac 5 2}$ & $\frac 1 {\sqrt{10}}$ 
 & $\frac 1 {\sqrt{10}}$ & 0.23 & 1.9 & 180 \\ 
 $Q+D$		& -$\frac 1 {6}$ &$\frac 1 {2}$ & 0 & 17  & 22 & 0 & $\frac 1 {3 \sqrt{2}}$ & $\frac 1 {\sqrt{2}}$ & $\frac 1 {\sqrt{2}}$ & 2.9 & 6.1 & 740 \\ 
 $U+E$		& -$\frac 5 {3\sqrt{6}}$ & 0 & $\frac 1 {2\sqrt{6}}$ & 0.57  & 0.08 & 120 & $\frac 7 {3 \sqrt{2}}$ & 0 & $\frac 1 {2\sqrt{2}}$ & 0.57 & 0.08 & 60 \\ 
 $G+E$		& -$\frac 4 3$ & 0 & $\frac 1 2$ & 0.57  & 0.08 & 180 & $\frac {\sqrt{2}}  3 $ & 0 & $\sqrt{2}$ & 0.57& 0.08 & 12000 \\ 
 $U+N$		& $\frac 4 {3\sqrt{6}}$ & 0 & $\frac 1 {2\sqrt{6}}$ & 0.57  & 0.08 & 180 & $\frac 4 {3\sqrt{2}}$ & 0 & $\frac 1 {2\sqrt{2}}$ & 0.57& 0.08 & 180 \\ 
\end{tabular}
\end{center}
\caption{\em Anomaly coefficients  for the $\eta$ and $\eta'$ singlets as predicted by a sample of vector-like confinement models. 
The $Q,U,D,L,E,G,\ldots$ fermionic vector-like multiplets are precisely defined in~\cite{strongDM}.}
\label{table:vectorlike}
\end{table}

Techni-eta (singlet techni-pions) have $\X F_{\mu\nu}\tilde F_{\mu\nu}$ interactions to SM vectors, as in eq. (\ref{eq:opsSU2}) but no coupling 
to SM fermions at leading order. The translation to the notation of eq. (\ref{LPanomaly}) is given by
\begin{equation}
\frac 1 {\tilde{\Lambda}_{i} }= \frac {c_{i}} {8\pi^2 f}  \qquad\hbox{where }  i = \{B,W,G,\gamma\},
\end{equation}
From this the widths into photons and gluons read
\begin{equation}
\frac {\Gamma_{\gamma\gamma}}M=c_\gamma^2\frac {\alpha^2}{64\pi^3} \frac {M^2}{f^2}\,, \qquad\frac {\Gamma_{gg}}M=c_G^2\,\frac {\alpha_3^2}{8\pi^3} \frac {M^2}{f^2}\,,
\label{eq:widthano}
\end{equation}
and similarly for other channels. The hypercharge, electroweak and QCD anomaly coefficients  are explicitly given by
\begin{equation}
c_B=2\NTC\,{\rm Tr} (T_S Y^2)\  ,\qquad c_W \delta^{ab}=2\NTC\,{\rm Tr}\, (T_S T^a T^b)  \qquad\hbox{and}\qquad 
c_G\delta^{AB}=2\NTC\, {\rm Tr} \,(T_S T^A T^B).
\end{equation}
Furthermore $c_\gamma =2\NTC\, \Tr( T_S Q^2)=c_B +c_W$. Here $T^a$ are the $\SU(2)_L$ generators
(which satisfy $\Tr(T^a T^b) =\delta^{ab} n(n^2-1)/12$ in the $n$-dimensional irreducible
representation), $T^A$ are the $\SU(3)_c$ generators, and $T_S$ is the chiral symmetry generator associated to $\X$. 
The anomaly coefficients are model-dependent numbers determined by the quantum numbers of the fermions under the SM.
The key novelty with respect to the perturbative models is that $1/f$ plays the role of $y_f/M_f$.
Strong interactions typically give $M/f \sim 4\pi /\sqrt{\NTC}$ for composite states or smaller if the scalar $\X$ is a PNGB.
This is effectively equivalent to allowing the Yukawa couplings introduced in the perturbative models of section~\ref{perturbative}
to become maximal (as large $g_* \sim 4\pi/\sqrt{\NTC}$ for QCD-like theories), thereby enhancing the $\X\to\gamma\gamma,gg$ widths. 
Notice that, since the anomaly coefficients are proportional to $\NTC$ while $M/f$ decrease with $1/\sqrt{\NTC}$,
the maximum width increases linearly 
with $\NTC$.

\medskip

Assuming that QCD processes dominate over electroweak ones,
$\Gamma(\X \to gg)\gg  \Gamma(\X \to \gamma\gamma,ZZ,WW,Z\gamma)$, 
and the absence of other decay channels,  $\Gamma_{\gamma\gamma}/M\approx 10^{-6}$ (see fig.\fig{figGG}a) 
in order to reproduce the signal rate. Comparing with eq. (\ref{eq:widthano}) we find,
\begin{equation}
\frac M f   \approx \frac {6}{c_\gamma} \,.
\end{equation}
Next, again assuming that the total width $\Gamma$ of $\X$ is dominated by decay into gluons we derive
\begin{equation}
\Gamma \approx  \Gamma_{gg} =\frac {8\alpha_s^2}{\alpha^2}\frac {c_G^2}{c_\gamma^2} \Gamma_{\gamma\gamma}\approx\frac {c_G^2}{c_\gamma^2}\,{\rm GeV}\,.
\label{eq:ratiomf}
\end{equation} 
As discussed in section 2 the $gg\to\X\to \gamma\gamma$ scenario without other decay channels cannot explain the total width of 45 GeV
favoured by ATLAS.
If allowed by phase space, $\X$ can also decay into lighter techni-pions (2 or 3, depending on whether the T-violating $\theta_{\rm TC}$ angle in the strong sector is different from zero). 
If present this new decay mode could be dominant and give a sizable width $\Gamma$ induced by strong interactions. 
$S$ could also decay into Higgs and lighter techni-pions if the quantum numbers of the new fermions allow for Yukawa couplings. 
In this more general setting the signal and width are reproduced for
\begin{equation}
\frac M f \approx \frac {14}{\sqrt{c_G\,c_\gamma}}\,\left(\frac{\Gamma}{45\GeV}\right)^{\frac 1 4}\,.
\end{equation}
For the central value of the width indicated by ATLAS we need $\Gamma_{\gamma \gamma}/M >   10^{-4}$ so that
\begin{equation}
\frac M f > \frac {50}{c_\gamma}.
\end{equation}
This equation implies that $M/f$ should be close to maximal unless the strong sector has large $\NTC$.
The latter option is preferred in order to avoid coloured states around $M$ but this could ruin the perturbativity 
of SM couplings at high energies.

\medskip

All the coefficients of the effective operators in eq. (\ref{eq:opsSU2}), and from here the 
$\X\to\gamma\gamma$, $\gamma Z$, $ZZ$, $WW,gg$ widths, are predicted in terms of the anomaly coefficients, 
as discussed in section~\ref{spin0}. The experimental limit on decays into massive SM vectors are satisfied if $-0.3 < c_W/c_B <2.4$.
If future data will see the excess in more than one final state channel, the ratios of cross sections will be predicted
in terms of the ratio of the corresponding anomaly coefficients, which form a discrete set
(unless multiple techni-$\eta$ are present and mix among them). 
The total width of the resonance determines in turn the width  in all the channels, including invisible.



\medskip

Examples of concrete models are listed in table~\ref{table:vectorlike}.
Let us discuss for instance the $Q\oplus D$ model: we extend the Standard Model by adding an $\SU(\NTC)$ gauge theory with fermions
$\Q =(\NTC,3,2)_{1/6}\oplus (\NTC,3,1)_{-1/3}$ plus their conjugates $\bar\Q$ under $\SU(\NTC)_c\otimes \SU(3)_c\otimes \SU(2)_L\otimes {\rm U}(1)_{Y}$.
This theory contains acceptable techni-baryon DM candidates and could achieve SM gauge couplings unification around $10^{17}\GeV$~\cite{strongDM}.
The techni-strong dynamics spontaneously breaks the accidental $\U(9)_L\otimes \U(9)_R$ global flavour symmetry 
down to the diagonal subgroup $\U(9)$.  This symmetry is explicitly broken by the SM gauge interactions, Yukawa interaction with the elementary Higgs 
 and  by the techni-fermion masses. The techni-pions decompose under the SM as,
\begin{equation}
 2 \times (1,1)_0+ 2 \times (8,1)_0 + (8,3)_0+ (1,3)_0+(1,2)_{\pm \frac 1 2}+ (8,2)_{\pm \frac 1 2}.
\end{equation}
The two singlet techni-$\eta$ do not receive mass from SM gauge interactions.
One of two singlets corresponds to the flavour generator proportional to the identity:
it is  the analog of the $\eta'$ in QCD and is anomalous under the new strong interactions, with mass given by eq.~(\ref{massone}).
The anomaly coefficients and partial widths are given in table \ref{table:vectorlike} and are compatible with the present experimental bounds
on $\Gamma(S\to Z \gamma)$ and $\Gamma(S\to ZZ)$ in table \ref{tabounds}.
The second singlet can be lighter but has a vanishing colour anomaly so that it cannot be produced through gluon fusion. 
This is however an accidental feature of this model, not shared by similar models. 
For example in the model with the unified representation $L+D$, the lighter singlet $\eta$ is viable and predicts widths in other channels observable
in the near future.  Due to the absence of Yukawa couplings in this model the specie numbers $L$ and $D$ are conserved so that techni-pions made 
of $LD$  (in the $(3,2)_{ 1/6}$ SM representation)  are long lived, unless non-renormalizable operators are added.
Other models are found in table \ref{table:vectorlike}.

\subsubsection{$\mathcal Q$-onium}

Another possibility appears if the extra vector-like fermions $\mathcal Q$ have masses which are comparable or slightly above the confinement scale $\Lambda$ (similarly to charm quarks in QCD) and  chiral symmetry is not spontaneously broken. The lightest composite states will be $\mathcal Q$-onium like and glueballs.
If $\mathcal Q$ have colour and electromagnetic charges, these $\mathcal Q$-onia can couple to $gg$ and $\gamma\gamma$ with tree-level strengths.
In simple potential models of $\mathcal Q  \bar{\mathcal Q}$ resonances, the corresponding decay widths are given by \cite{charmonium} 
\beq
\frac{\Gamma(\X \to \gamma \gamma)}M \sim  \alpha^2   Q_{\mathcal Q}^4 F_\X^2
 \,, \quad   \frac{\Gamma(\X\to gg)}{\Gamma(\X \to \gamma \gamma)} =  8 \frac {\alpha_3^2}{\alpha^2}\left[ \frac{I_{r_{\mathcal Q}}}{d_{r_{\mathcal Q}}Q_{\mathcal Q}^2}\right] ^2   \,,
\eeq
where $Q_{\mathcal Q}$ and $r_{\mathcal Q}$ are the charge and colour representation of ${\mathcal Q}$. Furthermore, for states with
angular momentum $J$,  $F_\X$ is the $J$-th radial derivative of the $\mathcal Q \bar{\mathcal Q}$ wavefunction at the origin, in units of $M$. For $\Lambda < 2 M_{\mathcal Q}$ it is expected to scale as $F_\X \sim (\Lambda/2M_{\mathcal Q})^{(J+3)/2}$. For $\Lambda \sim  2 M_{\mathcal Q}$ this could produce a large width in photons 
and decays into techni-glueballs (if  kinematically allowed) can account for the total width.
For example  the $c \bar{c}$ state $\eta_c[2980]$
in QCD has
$\Gamma(\eta_c \to \gamma \gamma)/M\approx 2 \times 10^{-6}$. This could be increased by lowering the fermion mass or increasing $\NTC$.

\section{Decays into Dark Matter?}\label{DM}
An interesting speculative possibility is that a dominant or substantial width of $S$ is into Dark Matter particles (see also~\cite{DM}).
This can be realised by identifying one of the $\Q$ particles in eq.\eq{YukSQ} with Dark Matter,
and assuming that it is a singlet under the SM gauge group.\footnote{A DM multiplet with electroweak interactions
could also generate $S\to\gamma\gamma$ at one loop.  
Unfortunately, in view of gauge annihilations, any electroweak multiplet would have a 
thermal relic abundance smaller than the DM relic abundance, if its mass is below $M/2$ (and above $M_W$, 
as suggested by experimental constraints)~\cite{MDM}.}
We consider the following models.
The new scalar $S$ couples to a single parton pair $\wp\bar\wp$ in the proton
with width $\Gamma_{\wp\wp}$, 
to photons with width $\Gamma_{\gamma\gamma}$,
and to DM with width  such that its total width is $\Gamma=0.06 M$, as favored by ATLAS.
We explore, in turn, all partons in the proton, $\wp=\{g,u,d,s,c,b\}$.
In the right panel of fig.~\fig{good} we show the regions in the $(\Gamma_{\wp\wp},\Gamma_{\gamma\gamma})/M$ plane
that reproduce the $pp\to S\to\gamma\gamma$ cross section favored by LHC,
while satisfying all bounds, in particular from dijet and invisible final states.

{Neglecting the possibility that the width is dominated by the $\gamma\gamma$ channel,} we see that a dominant decay width into DM is needed if $S$ is produced through the $\wp\in\{g,u,d\}$ partons, while 
the decay width into DM can be small or even vanishing if instead $S$ only couples to $\wp\in\{s,c,b\}$.
In all cases a relatively large $\Gamma_{\gamma\gamma}/M \circa{>}10^{-5}$ is needed.
SM loops alone produce a much smaller effect, so that extra charged states must be added.
Even so, reproducing such a large rate needs uncomfortably large charges and/or couplings.
We thereby focus on the smallest allowed values of $\Gamma_{\gamma\gamma}$, 
which correspond to values of the $S$ couplings to partons around the upper bounds listed
in the following table:
$$
\begin{array}{c|c}
\hbox{parton}&\hbox{Upper bound} \\
\wp\wp  & \hbox{on the partonic coupling}\\ \hline
gg &    M/\Lambda_g, M/\tilde\Lambda_g <0.05 \\
u\bar u & y_{uS},\tilde y_{uS}<0.08\\
d\bar d  &y_{dS},\tilde y_{dS}<0.11 \\
s\bar s  &y_{sS},\tilde y_{sS}<0.68\\
c\bar c  &y_{cS},\tilde y_{cS}<0.71\\
b\bar b & y_{bS},\tilde y_{bS}<0.71
\end{array}$$

\subsection{Cosmological Dark Matter abundance}
We next describe how the DM thermal relic abundance can be computed in a quasi-model-independent way.
In the usual relic abundance computation, the key quantity is the thermal average of
the annihilation cross section of two DM particles, which depends on $s=(p_1 + p_2)^2$
where $p_1$ and $p_2$ are quadri-momenta of the DM particles.
As usual, DM annihilation freezes out when DM is non relativistic,
at a temperature ${T_f}\approx {\MDM}/{25}$.
In most of the parameter space, the DM annihilation cross section can then be expanded in the non-relativistic limit\footnote{In our case, the above approximation fails only in the narrow range  $0\le M_S - 2\MDM \circa{<} T_f$, where
the thermal average of the cross section receives significant contributions from configurations
where the DM particles can acquire enough energy to perform a resonant scattering.
The width of the left-handed size of the narrow dip is always a few $\%$.} in powers of the
relative DM velocity 
$v=2\sqrt{1-4\MDM^2/s}$ as
\beq \sigma v =  \sigma_0 + v^2 \sigma_1 + {\cal O}(v^4).\eeq
Then, the approximated solution to Boltzmann equations dictates that the 
DM thermal relic abundance reproduces the observed cosmological DM abundance when
\beq \sigma_0 + \frac{3T_f}{\MDM} \sigma_1\approx
\frac{1}{(22\TeV)^2}  .\eeq
The sub-leading $p$-wave term is accurate only when the leading $s$-wave term vanishes.

\medskip

\begin{figure}[t!]
\centering
$$\includegraphics[width=0.45\textwidth]{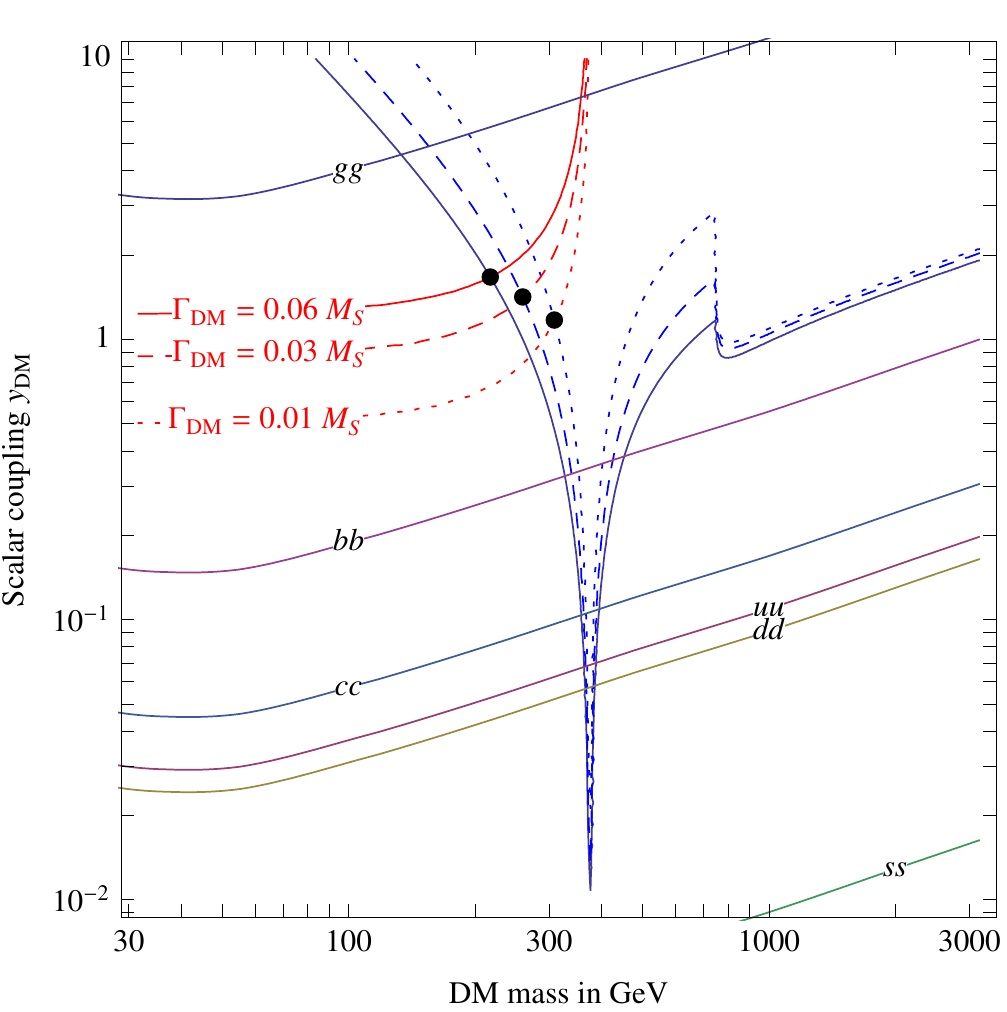}\qquad
\includegraphics[width=0.45\textwidth]{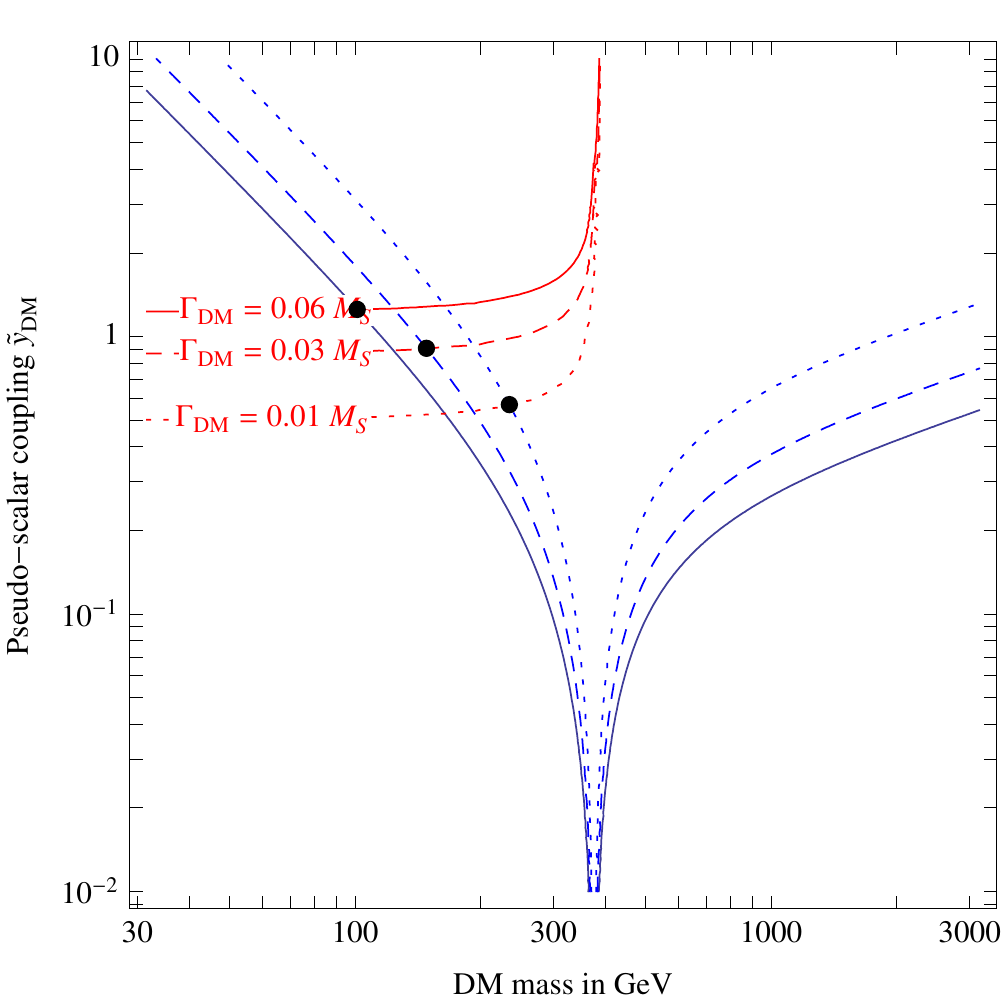}$$
\caption{\em Regions in the plane (DM mass, DM coupling to $S$) where
DM has the observed relic abundance (bounded by blue lines), assuming 3 values of the $S$ width: $\Gamma/M = 0.06$ (solid), 
$0.03$ (dashed), $0.01$ (dotted).
Furthermore, the red contour lines show when such width is reproduced dominantly through $S$ decays into DM.
The dots indicate the points where both conditions are satisfied, as described in table~\ref{tab:MDM}.
The contours labeled by parton pairs show the upper bounds from spin-independent DM direct detection
assuming maximal couplings of $S$ to partons.
Such bounds are negligible in the pseudo-scalar case.
\label{fig:DMres}} 
\end{figure}

\begin{table}
$$\begin{array}{c|cc|cc}
& \multicolumn{2}{c|}{\hbox{scalar coupling}}& \multicolumn{2}{c}{\hbox{pseudo-scalar coupling}}\\
\Gamma/M & \MDM &y_{\rm DM} & \MDM & \tilde y_{\rm DM} \\[0.2mm]  \hline
0.06& 217\GeV & 1.67 & 101 \GeV& 1.25\\
0.03 & 260\GeV & 1.41 & 148 \GeV& 0.91\\
0.01 & 308\GeV & 1.17 & 234\GeV & 0.57
\end{array}$$
\caption{\em Predictions for the DM mass and the DM coupling to $S$.\label{tab:MDM}}
\end{table}

A $2\to 2$ cross section is given by the usual general formula
\beq \sigma = \frac{p_{\rm out}}{p_{\rm in}} \frac{\sum_{\rm out}\langle  | \mathscr{A} |^2 \rangle_{\rm in}}{16\pi s}\eeq
where $p_{\rm in(out)}$ is the momentum of an in-coming (out-going) particle in the center-of-mass frame.
The only DM annihilation process present for $\MDM < M_S$ is the $s$-channel diagram where a virtual $S^*$ is exchanged
between DM and a generic set of SM final-state particles.
Then, the amplitude for this process can be decomposed as $\mathscr{A}= \mathscr{A}_{\rm in} \mathscr{A}_{\rm out}/\Pi$ where 
the resummed propagator is written in terms of the two-point function as
$\Pi(s)= s-M_S^2 +\delta \Pi$.
The loop corrections $\delta \Pi$ can be complex describing the width of a virtual $S^*$.
In order to compute the thermal relic abundance we need $\sigma v$, where $v$ is the relative DM velocity, $v = 2p_{\rm in}/\MDM$:
\beq \sigma v = \frac{ \langle|\mathscr{A}_{\rm in}|^2\rangle}{M_{\rm DM}} \times \frac{ 1}{|\Pi |^2}  \times \sum_{\rm out} \frac{ |\mathscr{A}_{\rm out}|^2 p_{\rm out}}{8\pi s} .
\eeq
In the above equation:
\begin{itemize}
\item The latter factor is $\Gamma(S^*\to \hbox{SM})$, the decay width of a virtual $S$ with mass $\sqrt{s}\approx 2\MDM$ into SM particles.
\item The factor in the middle gives the Breit-Wigner shape, with a generalised off-shell width.
Its imaginary part $\sqrt{s} \Gamma_{S^*}$ can be approximated as $M_S \Gamma_S$, given that it is relevant only when $S$ is almost on-shell.
\item
The first factor has been renamed as
\beq \Gamma_v( \hbox{DM}\to S^*)\equiv   \frac{ \langle|\mathscr{A}_{\rm in}|^2\rangle}{\MDM}.\eeq
In view of the average over initial DM states, it does not increase if multiple DM states are introduced.
\end{itemize}
We then obtain the final formula for the $s$-channel DM annihilation cross section at $s\circa{>} 4\MDM^2$:
\beq\label{eq:sigmavres}
\sigma v = \frac{\Gamma_v(\hbox{DM}\to S^*)
\Gamma(S^*\to \hbox{SM})}{(s-M^2)^2 + M^2 \Gamma^2}.
\eeq
Let assume, for example, that DM is a Dirac fermion.  
The factor $\Gamma_v$, expanded in the non-relativistic limit, is
\beq \Gamma_v(\hbox{DM}\to S^*) = 2\MDM \tilde{y}_{\rm DM}^2 +\frac{v^2}{2} \MDM y_{\rm DM}^2+\cdots\eeq
These are the factors that enter into the resonant DM annihilation cross section of eq.\eq{sigmavres}.\footnote{We also add the model-dependent 
$\bar \Psi_{\rm DM}\Psi_{\rm DM}\to SS$ scattering process, which only contributes
if $r=\MDM/M\ge 1$. 
Ignoring possible $S$ self-interactions, we find the following extra contribution to the DM annihilation cross section
induced by the couplings $y_{\rm DM}$, $\tilde y_{\rm DM}$.
The $s$-wave term is present only when CP is violated
\beq\sigma_0 =  \frac{ y_{\rm DM}^2 \tilde y_{\rm DM}^2}{2\pi \MDM^2}\frac{r^3 \sqrt{r^2-1}}{(2r^2-1)^2}.\eeq
The $p$-wave term is
\beq \sigma_1 =  \frac{1}{24\pi\MDM^2}\frac{r^3\sqrt{r^2-1}}{(2r^2-1)^4}\bigg[
y_{\rm DM}^4   (2-8r+9r^4)+ 
\tilde y_{\rm DM}^4 (r^2-1)^2-
3y_{\rm DM}^2 \tilde y_{\rm DM}^2 \frac{1-8r^2+20r^4+12 r^6}{2(r^2-1)}
\bigg].\eeq}
Fig.~\ref{fig:DMres} shows the regions in the (DM mass, DM coupling) plane where the DM cosmological thermal abundance is reproduced
 (in blue) and where the total width is reproduced in terms of decays into DM, with width
\beq
\Gamma(S\to \bar\Psi_{\rm DM}\Psi_{\rm DM}) = \frac{M_S}{8\pi} \Re \bigg[
\tilde y_{\rm DM}^2  \bigg(1-\frac{4\MDM^2}{M_S^2}\bigg)^{1/2}
+ y_{\rm DM}^2 \bigg(1-\frac{4\MDM^2}{M_S^2}\bigg)^{3/2}\bigg].
 \eeq 
Table~\ref{tab:MDM} lists the values of the DM mass and couplings such that both conditions are satisfied.
From figure~\ref{fig:DMres} one can easily derive the analogous results in presence of multiple DM states or with a different total width of $S$.

\subsection{Direct detection}

Given that $S$ couples as $S(J_{\rm SM}+J_{\rm DM})$ (where the factors $J$ have been defined previously), 
by integrating out $S$ one obtains the effective Lagrangian relevant for direct detection of Dark Matter, $\Lag_{\rm eff} = J_{\rm SM} J_{\rm DM}/M^2$.
The dominant Spin-Independent direct detection cross section is produced only 
if $S$ is a scalar that couples to the SM particles through the $\Lambda_g$ or $y_{Sq}$ couplings and to DM
thought the $y_{\rm DM}$ coupling.
The Spin-Independent direct detection cross section is~\cite{Panci}
\beq\sigma_{\rm SI}=\frac{m_N^4 y_{\rm DM}^2}{\pi M_S^4 }\bigg[ -  \frac{12\pi^2 f_g }{9\Lambda_g} +
\sum_q\frac{y_{qS} f_q }{y_q v}\bigg]^2 \label{eq:sigmaSI}.
 \eeq
The protonic matrix elements are
$f_u \approx 0.019$, $f_d\approx 0.024$, $f_s\approx 0.093$, $f_g\approx f_c\approx f_b\approx f_t \approx \frac{2}{27}(1-f_u-f_d-f_s)\approx 0.064$.
Furthermore $y_q$ are the usual Yukawa couplings of quarks $q$ to the SM Higgs doublet
with vacuum expectation value $v=174\GeV$.
Inserting numbers, eq.\eq{sigmaSI} becomes
\beq\sigma_{\rm SI}= 4~10^{-47}\cm^2 \,y_{\rm DM}^2 \bigg[-27. \frac{M_S}{\Lambda_g}
+0.30 \frac{y_{uS}}{y_{u}}+0.38\frac{y_{dS}}{y_{d}}+1.5 \frac{y_{sS}}{y_{s}}+
\frac{y_{cS}}{y_{c}}+\frac{y_{bS}}{y_{b}}+\frac{y_{tS}}{y_{t}}\bigg]^2.\label{eq:sigmaSIN}\eeq
If $S$ couples to gluons, the bound $y_{\rm DM}\circa{<}4$ (for $\MDM\approx 100\GeV$) is weak, given that  $M/\Lambda_g <0.05$.
Stronger bounds are obtained if $S$ couples only to quarks:
the bounds on $y_{\rm DM}$ are shown in the left panel of fig.~\ref{fig:DMres},
where we consider $S$ coupled to each parton $\wp$ in turn,
and in each case we assume the maximal value of the coupling of $S$ to the parton ($M_S/\Lambda$ or $y_{qS}$),
in order to reduce $\Gamma_{\gamma\gamma}$.

The previous bounds hold for the case of scalar couplings.
Bounds on $\tilde y_{\rm DM}$ are weaker by orders of magnitude.
Similarly, bounds on $y_{\rm DM}$ and on $\tilde y_{\rm DM}$ are much weaker if
$S$ is a pseudo-scalar state that couples through $\tilde\Lambda_g$, $\tilde y_q$.

Indirect detection bounds are instead negligible for the case of a scalar $S$, given that 
it gives a $p$-wave suppressed DM annihilation cross section.

\begin{figure}[t]
\begin{center}
$$ \includegraphics[width=0.6\textwidth]{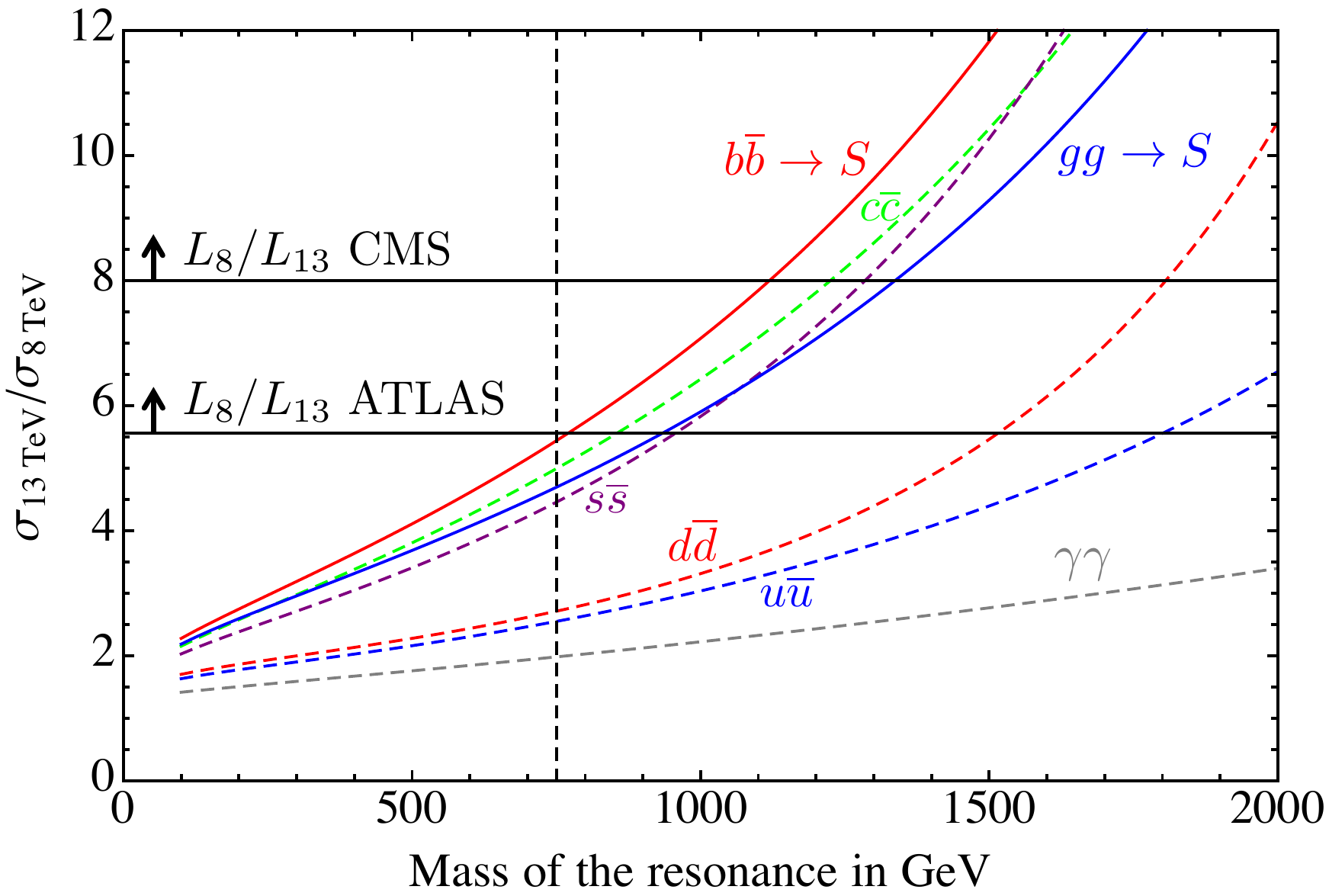}$$
\caption{\em 
 Ratio of $pp$ cross sections at $\sqrt{s}=13\TeV$ and $8\TeV$
 for producing a narrow resonance $\X$ with mass $M$  computed for different initial partons,
compared to the inverse ratio of luminosities accumulated by CMS (upper) and ATLAS (lower).  This reflects the relationship of the total number of events observed between $8$ and $13 \TeV$, but does not reflect the significance, which depends additionally on the background at the two energies.
\label{fig:figratio}}
\end{center}
 \end{figure}
 
 \section{Improved Run 1/Run 2 compatibility from a parent resonance}\label{heavier}
While the simplest interpretation of the diphoton peak  is a resonance with mass $M=\Excess$, a closer inspection of the data may suggest to consider more complicated kinematics.  Specifically, the $\Excess$ resonance is not apparent in LHC data at 8 TeV, save for a slight upward fluctuation.  In eq.\eq{sigma813} we see that the
8 TeV data prefer a cross section which is $0.06\pm0.06$ times smaller than that at 13 TeV, 
and a naive fit of the $pp\to\gamma\gamma$ rates in eq.\eq{sigma813} shows that the data at $\sqrt{s}=8\TeV$
are incompatible with those at $\sqrt{s}=13\TeV$ at $95\%$ confidence level if the
cross section grows less than about a factor of 3.5.

Figure\fig{figratio} shows how the cross section for producing a resonance with mass $M$ increases depending on the initial partons.
The cross section for producing a resonance at $M=\Excess$ increases by a factor $4.7$ for $gg$ initial partons and by $5.4$ for $b\bar b$ partons.
This roughly compensates for the reduced  luminosity accumulated during Run 2 (3.6 fb$^{-1}$ in ATLAS and 2.6 fb$^{-1}$ in CMS) with respect to Run 1 (about 20 fb$^{-1}$).
Furthermore, the SM $\gamma\gamma$ background increases by a smaller factor $\approx 2.3$.
Indeed it is dominated by $q\bar q\to\gamma\gamma$ and
$\sigma(pp\to\gamma\gamma)\approx 6\, \fb$ at 8 TeV and $\approx14\,\fb$ at 13 TeV, after imposing  $m_{\gamma\gamma}>\Excess$ and standard cuts.
We thereby see that Run 2 data can already be more powerful than Run 1 data.  A stronger increase of the signal/background ratio is obtained if the $750$ GeV diphoton resonance originates from the decay of a heavier resonance, according to the process depicted in fig.\fig{diagram}.  This scenario could arise in both perturbative or strongly coupled models.  Here we will just consider the generic predictions.

\subsection{General framework}
The basic framework is that a heavy ``parent'' resonance $P$ is produced at the LHC, $pp \to P$.  Then $P$ decays to a $750$ GeV resonance $S$ and another state $R$.  Finally $R$ decays to final state particles which evade detection.  If they are dark matter particles we will denote them by $\chi$.  Alternatively, $R$ may also cascade through hidden sector states terminating in a large multiplicity of soft hadrons, as may occur in `Hidden Valley' scenarios \cite{Strassler:2006im}.  As these additional states evade detection the only observed end product is $S$, which decays to two photons.  Let us first consider the case that $\chi$ is a stable dark matter particle.  
\begin{figure}[t]
\begin{center}
$$ \includegraphics[width=0.35\textwidth]{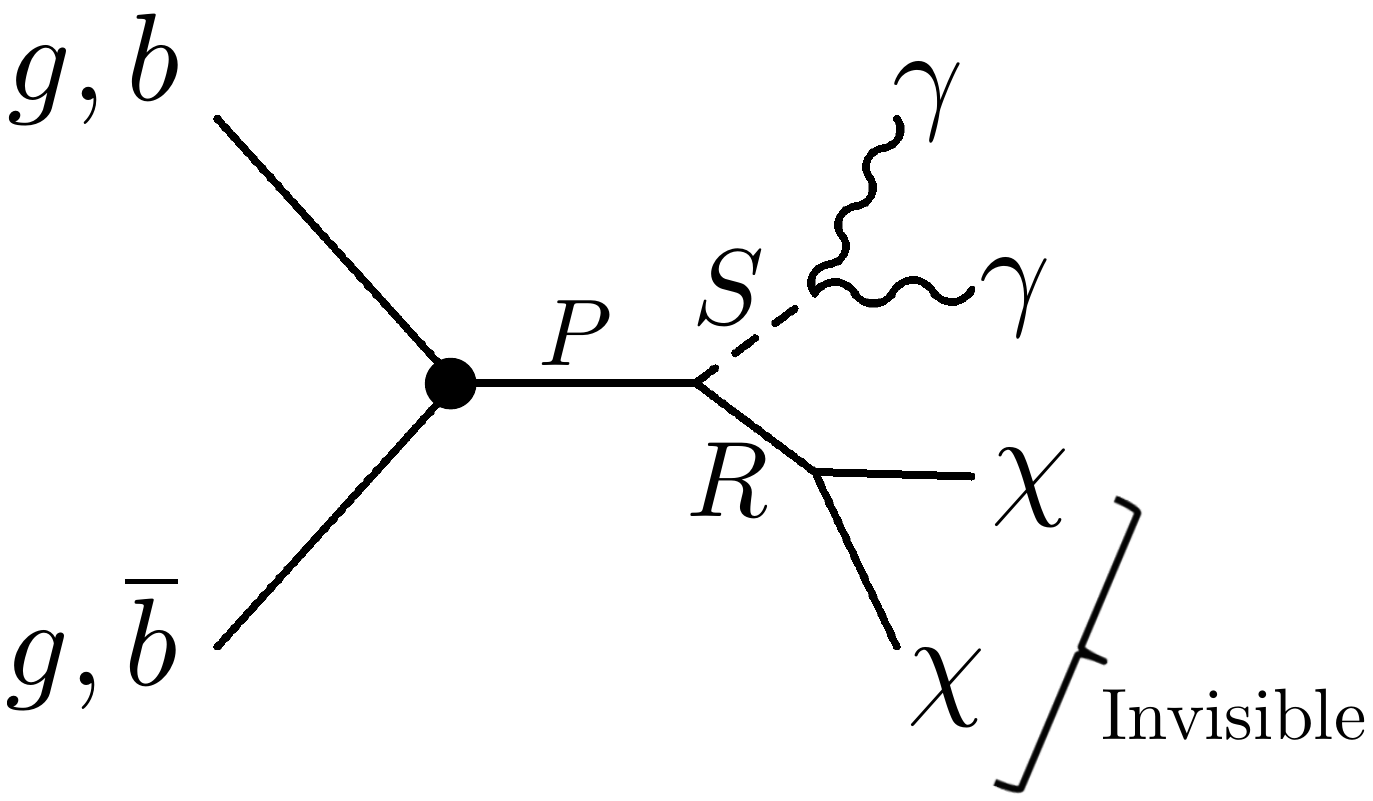} \qquad \includegraphics[width=0.35\textwidth]{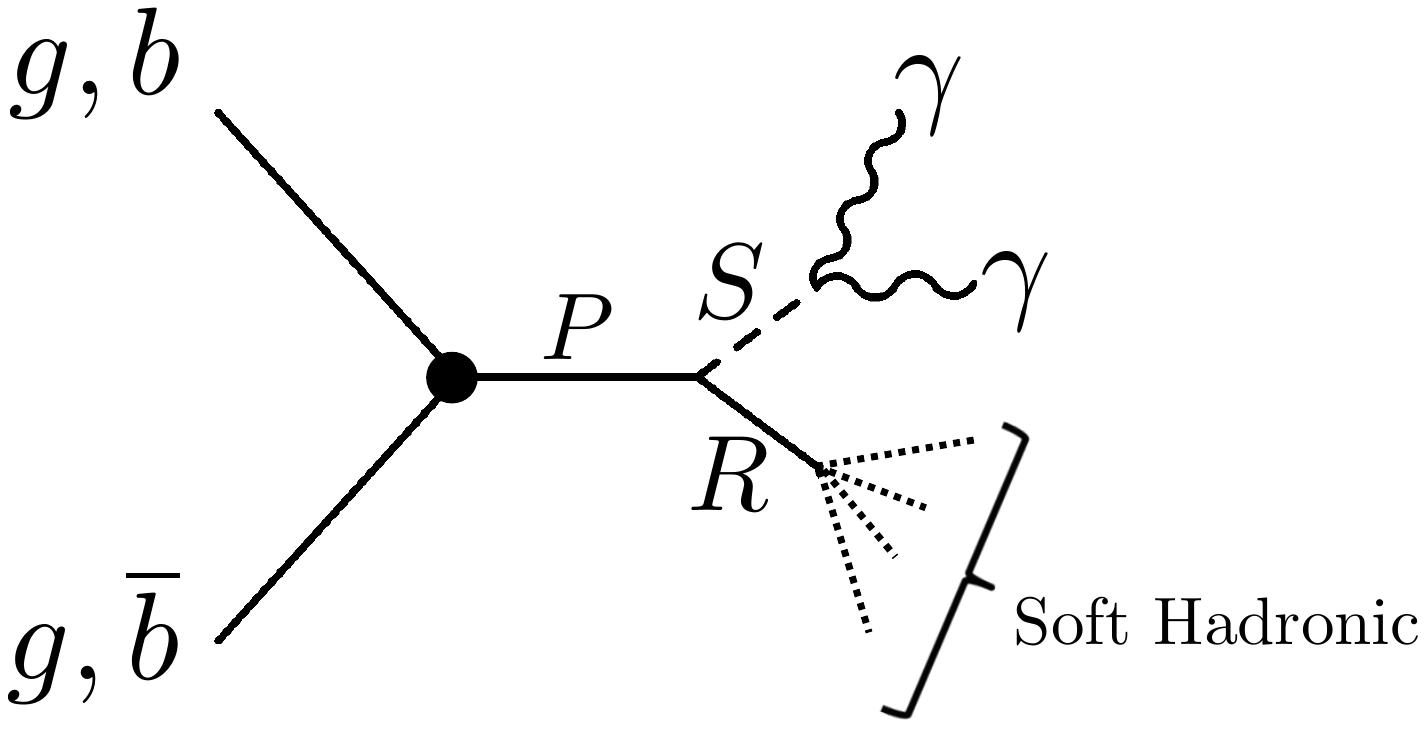}$$
\caption{\em 
 Production of a diphoton resonance $S$ from the decay of a heavier ``parent''  resonance $P$.  The additional decay products $\chi$, originating from the decay of $R$, may be stable dark matter candidates and escape undetected (left) or may decay into high multiplicity soft hadronic final states (right), which are difficult to discriminate from soft QCD processes.  For the case of dark matter production significant missing energy is avoided by a small mass splitting $m_P = m_S + m_R+ \Delta$, where $\Delta \sim 10 \GeV$ is small enough to suppress the missing energy signature. 
\label{fig:diagram}}
\end{center}
 \end{figure}

There are a number of possibilities for the nature of the particles involved, which we will now discuss.
\begin{itemize}
\item  For a $gg$ initial state a scalar parent resonance $P$ could be produced via a loop of heavy particles, denoted with a dot in fig.\fig{diagram}.  This scalar resonance could decay to two scalars, $S$ and $R$.
\item  For a $b\overline{b}$ initial state $P$ could be a heavy vector boson, which decays to a scalar $S$ and another vector $R$.  For the remainder of this work we will consider the $gg$ initial state and scalar $P$, $S$, and $R$.
\item  The most minimal possibility for a $gg$ initial state would be if $P$ decays to two $S$ resonances (i.e.\ $R=S$ in fig.\fig{diagram}).  In this way, if $S$ has the largest branching ratio to dark matter $S\to \chi \chi$ and a smaller branching ratio to diphotons $S\to \gamma \gamma$, then the majority of the observed events would be in the $gg \to P \to S S \to \chi \chi \gamma \gamma$ final state.  Eventually an observation of a pair of diphoton resonances would be expected from $gg \to P \to S S \to  \gamma \gamma \gamma \gamma$, however this would depend on the diphoton branching ratio.  To make this setup even more appealing, it could be that the dark matter $\chi$ is in an electroweak multiplet, such that direct decays to dark matter generate the required width, and loops of charged dark matter partners generate the coupling to photons.  In this case final states $gg \to P \to S S \to \chi^+ \chi^- \gamma \gamma$ could occur, where $\chi^\pm \to \chi^0 + \pi^\pm$, where the final state pions are very soft.\footnote{Also for higher EW multiplets longer cascades could occur i.e.\ $S\to \chi^{+++} \chi^{---} \to \chi^{++} \chi^{--} \pi^+ \pi^- \to \chi^+ \chi^- 2 \pi^+ 2 \pi^- \to 2 \chi^0 3 \pi^+ 3 \pi^-$.}
\item  Finally, the missing energy spectrum can be significantly softened if the parent resonance decays immediately to a three-body final state, $P \to S R T$, where now $T$ is some additional state.  If $R$ and $T$, or their decay products, are invisible then a similar signature arises if $m_P = m_S + m_R+ m_T+ \Delta$, where in this case it is likely that $\Delta$ could be larger than for the two body decays while still suppressing the missing energy signature.
\end{itemize}

In the recently observed diphoton excess events there is no significant missing energy component, thus an immediate question in this setup is how the missing energy is hidden.  The only observable is the diphoton system, and all of the observed missing energy comes about through the boost of the diphoton system, which is proportional to the boost of $S$.  Since at leading order $P$ is produced at rest in the transverse direction, all of the transverse boost of $S$ comes from the decay $P \to R S$.  Thus the boost of $S$ may be suppressed if the mass splitting $\Delta = m_P-m_S - m_R$ is small.  We assume this small mass splitting comes about accidentally, although it would be interesting to see how it may be motivated within a full model.  This requires a coincidence of masses, however if such a coincidence is tolerated then this setup may realise an observed $750$ GeV diphoton resonance which originates from the production of a much heavier parent resonance.

In the limit of a small mass splitting $\Delta$ the total momentum of the diphoton system ($p_S$) in the rest frame of the parent will be given by
\be
p_S = \sqrt{\frac{2 m_R m_S \Delta}{m_R + m_S}} ~~,
\label{eq:momentum}
\ee
thus, if $S$ is produced isotropically the missing energy spectrum will be approximately described by the following transverse momentum ($p_T$) distribution
\be
\frac{dN}{d p_{T}} = N \frac{p_T}{p_S \sqrt{p_S^2 - p_T^2}} ~~,
\label{eq:boost}
\ee
which is sharply peaked at values $p_T \sim p_S$.  

\begin{figure}[t]
\begin{center}
$$ \includegraphics[width=0.5\textwidth]{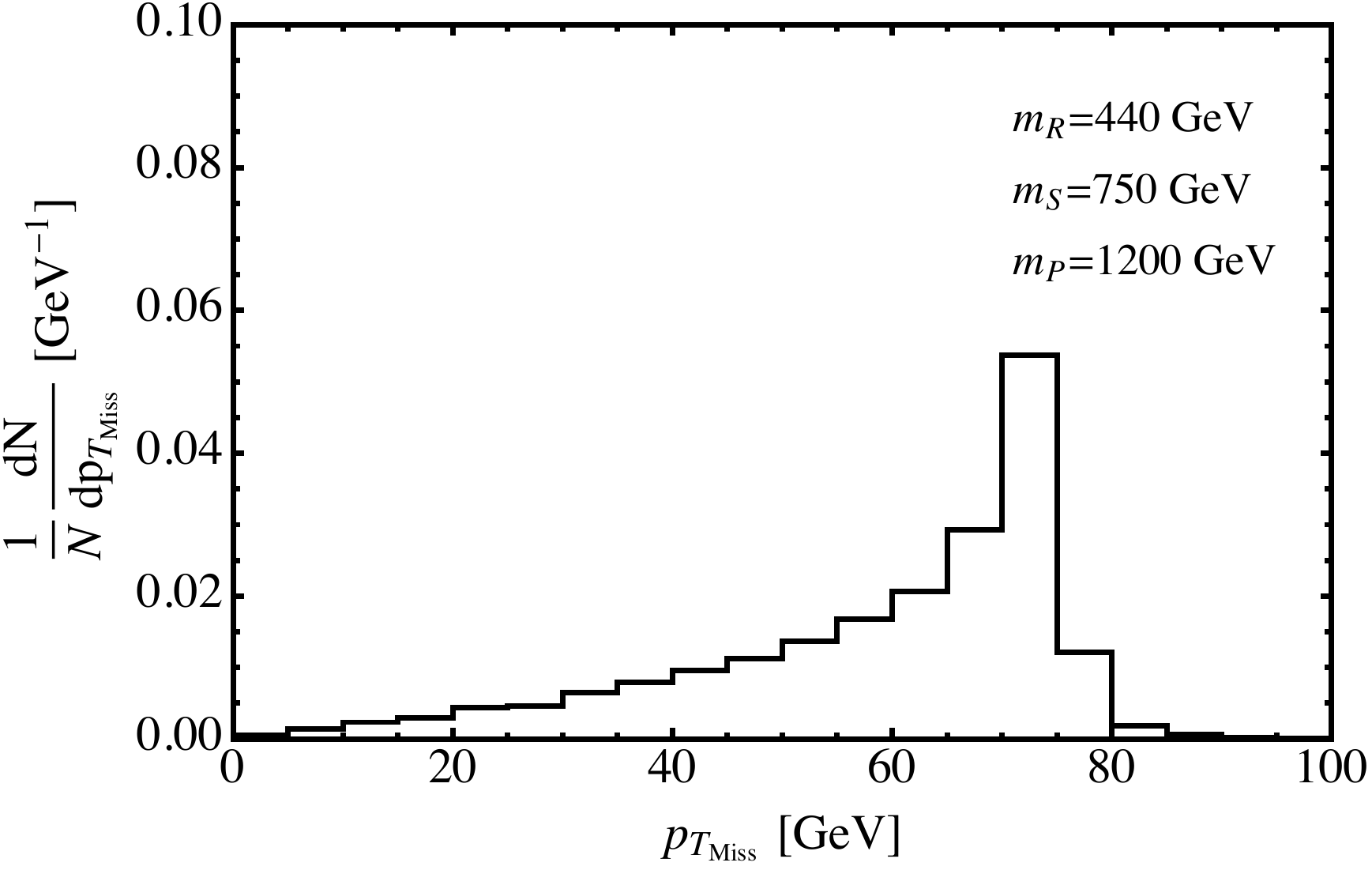}$$
\caption{\em 
Missing energy spectrum for the benchmark parameter point described in the text when $R$ decays to dark matter.  For $R$ decays to high multiplicity hadronic final states there would be no overall missing energy; however the diphoton system would have the same transverse boost.
\label{fig:figmet}}
\end{center}
 \end{figure}

\subsection{A benchmark model}
We will consider a model with a scalar parent resonance $P$ coupled to gluons as in eq.\eq{ops}.  The 750 GeV resonance $S$ will be coupled to photons as in eq.\eq{ops} and we will assume the width $\Gamma_S/m_S \sim 0.06$.  This could come about from any of the mechanisms discussed in sections~\ref{perturbative} and \ref{nonpert}.  The final ingredient is that $P$ will be coupled to $S$ and an additional state $R$ as
\be
V = \lambda m_P P S R ~,
\ee
where we have chosen to normalise the dimensionful coupling in units of the parent resonance mass.  A full phenomenological study of this scenario is beyond the scope of this work, however we will provide a benchmark scenario and use this to highlight the phenomenology.  From fig.\fig{figratio} we see that the cross section increase from 8 to 13 TeV for a gluon-initiated resonance approaches a factor of $7$ for a heavy resonance of mass $m_P \approx 1.2$ TeV (which is preferable over a factor $4.5$ for a $750$ GeV resonance).  With this in mind we take a benchmark set of masses
\be
m_P = 1.2 \text{ TeV} ,\qquad m_S = 750 \text{ GeV} ,\qquad m_R = 440 \text{ GeV}
\label{eq:bench1}
\ee
and benchmark couplings of
\be
\Lambda_{g,P} = 40 \text{ TeV},\qquad \lambda = 1.
\ee
To study the phenomenology we have implemented this model in MadGraph \cite{Alwall:2014hca}.  Events were showered in Pythia \cite{Sjostrand:2006za}.  Resonances are decayed using MadSpin \cite{Artoisenet:2012st}.  For these benchmark parameters the cross sections and branching ratios are
\be
\sigma(pp \to P) = 74 \text{ fb}, \qquad \text{BR}({P \to R S}) = 0.94, \qquad \text{BR}(P \to g g) = 0.06
\label{eq:bench2}
\ee
consistent with bounds on dijet resonances.  Thus, to achieve the needed total effective cross section for the diphoton resonances we require a branching ratio into photons of $\text{BR}(S \to \gamma \gamma) \approx 0.1$.  If the experimental hint for a width of $\Gamma_S \sim 45$ GeV was also to be generated then it would likely be required to generate a larger production cross section, in order to tolerate a smaller diphoton branching ratio.  This is because it would typically be difficult, especially in perturbative models, to have $\Gamma_{S\to \gamma \gamma} \sim 4.5$ GeV for a branching ratio $\text{BR}(S \to \gamma \gamma) \approx 0.1$.  More generally, the width and branching ratio parameters could be achieved by using the scenarios described in sect.~\ref{perturbative}.  In fig.\fig{figmet} we show the missing energy distribution of the events.  As can be seen, the majority of the events are at missing energies of $p_T \sim 70$ GeV.

\subsection{Suppressed parent resonance decays}
If the mass splitting $\Delta$ is small, the phase space for $P \to RS$ decays is greatly suppressed.  This can be seen from the fact that the width is proportional to $p_S/m_P$, where $p_S$ is given in eq.\eq{momentum}.  As $P$ has been produced in parton collisions at the LHC it may also decay into SM final states, thus for this class of models it may be necessary to ensure the $P \to RS$ decays to dominate.  We may understand this better by considering the benchmark model of the previous section.  The ratio of partial widths to gluons and hidden sector states is
\be
\frac{\Gamma(P \to g g)}{\Gamma(P \to R S)} = (8 \pi \alpha_3)^2 \frac{m_P^3}{\lambda^2 \Lambda_{g,M}^2} \sqrt{\frac{m_R+m_S}{2 m_R m_S \Delta}}  ~~.
\ee
Inserting the benchmark parameter values from eq.s\eq{bench1} and \eq{bench2} we see that this ratio is indeed small ($6 \%$) thus the great majority of $P$ production events at the LHC will result in $P\to R S$ decays.

\subsection{Possible signatures}
We see from fig.\fig{figmet} that in the future, if a scenario involving dark matter production is responsible for the observed diphoton excess, some events with missing energy in addition to the diphoton resonance should arise.  In particular, an ISR jet could boost the entire system of particles, thus a much larger missing energy signature may arise, depending on the model, in a monojet or monojet+diphoton search.  On the other hand, it is also notable that if $R$ is hidden by decays to high multiplicity soft hadronic final states rather than dark matter then, since $R$ is produced almost at rest in the transverse direction, this soft radiation would be distributed almost isotropically.  As the multiplicity would need to be high to hide the mass energy of $R$, the photon isolation from soft QCD radiation would likely be degraded.  Also, in this class of models, a dijet resonance at higher masses could possibly be observed from $gg \to P \to gg$ production, however this may be suppressed if $P$ decays dominantly via $P \to RS$.

There are more exotic possibilities concerning the embedding of this general framework into a complete model.  One which may address the question of the large width is, in the same spirit of sect.~\ref{many}, that the dark sector be composed of a multitude of states $S_1, S_2,..,S_N$ all with similar masses split by small amounts of $\mathcal{O}(\text{few GeV})$. This is natural if these states are in a multiplet of some approximate global symmetry, as with {\it e.g.} $\pi^0, \pi^\pm$.  If each of these states can decay as $S_i \to \gamma \gamma, \chi \chi$, where $\chi$ is the dark matter or some other invisible state, then if $\Gamma(S_i\to \chi \chi) \gg \Gamma(S_i\to \gamma \gamma)$ the dominant observable signature will come from $pp \to P \to S_i (\to \chi\chi) S_j (\to\gamma \gamma)$.  The diphoton invariant mass will be equal to $m_{S_j}$ thus a spectrum of diphoton resonances centered around some mean value, in this case $750$ GeV, would be observed.  The width of the total spectrum of diphoton lines may reproduce the required value of $\sim 45$ GeV.  In this way, measuring the observed diphoton excess would correspond performing dark sector spectroscopy.

In all of these scenarios the diphoton system would on average be boosted with a transverse momentum distributed according to eq.\eq{boost}, which is a characteristic signature of this setup.  For the benchmark we chose a typical boost of $\sim 70$ GeV, however this boost is essentially a free parameter and could take a range of values.  With this in mind it would be interesting to scrutinise the data to see if the diphoton system is on average boosted in the observed events.

\section{Conclusions}\label{section}
An excess in $pp\to \gamma\gamma$ events at $\sqrt{s}=13\TeV$ 
peaked at the invariant mass $M\approx \Excess$ has been claimed by the 
 ATLAS and CMS collaborations,  with a local statistical significance of $3.9$ and $2.6$ standard deviations respectively.
 The ATLAS data favour a relatively large width, $\Gamma/M\approx\GM$, a feature not apparent in the CMS data  
 which however have a smaller integrated luminosity.
 
The excess is not incompatible  with the $pp\to\gamma\gamma$
data taken at $\sqrt{s}=8\TeV$, provided that the $\Excess$ resonance production is initiated by
$gg$ or $b\bar b$ or $c\bar c$ or $s\bar s$ partonic collisions.
The compatibility of Run 1 and Run 2 data can be improved by assuming that the $\Excess$ resonance is produced though decays of a heavier `parent' particle. In section~\ref{heavier}
we propose a way to realise this process without extra hard particles or missing transverse energy in the event, as suggested by data.  The mechanism relies on a mild coincidence between the masses of different particles.  This scenario may also be connected to dark matter production at the LHC.

In section~\ref{pheno} we analyse the phenomenological properties of the simplest interpretation --- a bosonic $s$-channel resonance $\X$. The particle can have spin 0 or 2.
We extract from data its partial decay widths into the various SM channels.
A total width $\Gamma\approx \GM \, M$ is added as an optional constraint, given that this experimental information is 
particularly uncertain.
We also take into account the bounds from 8 TeV data on hypothetical resonances in
other channels, as summarised in table~\ref{tabounds}.

A particle coupled to gluons and photons (for example through loops of extra fermions) would be a simple theoretical option.
We find that $gg \to \X\to \gamma\gamma$ can fit the signal rate. Reproducing the total width, compatibly with the absence of dijet peaks, requires the presence of additional decay modes beyond $\X\to\gamma\gamma, gg$.
In order to avoid over-production of $\X$, most of its large width must be attributed to particles with small partonic   content of the proton.
At the same time these particles must make a final state not subject to strong experimental constraints.
We find that $t$, $b$ quarks are good candidates.
A decay width into extra invisible states, such as dark matter, is also a viable option.
Fig.\fig{good} shows how the various possibilities can reproduce the data.

In section~\ref{perturbative} we try to build weakly-coupled models that reproduce the required $\X$ partial widths. 
This can be easily done only if the constraint on the total width is ignored or reinterpreted. Indeed,
in section~\ref{many} we have shown how a multiplet of quasi-degenerate narrow resonances can fake the apparently large `width'.
This situation is automatically achieved in the case of a new heavy Higgs doublet, where the two neutral components can be split by the right amount through electroweak breaking effects.

If instead a genuine wide width is confirmed by data, a large $\Gamma(\X\to\gamma\gamma)$ is required
in order to  keep $\hbox{BR}(\X\to\gamma\gamma)$ at the value necessary to reproduce the rate of the observed excess. 
That  requires a sizeable coupling of $\X$ to two photons, which under broad circumstances implies a large number $N$ of degrees of freedom,
with possibly large gauge quantum numbers. 
We have shown that some coupling becomes non-perturbative not much above the TeV scale.
Such a situation normally arises in strongly coupled models.  

The hierarchy problem and its resolution by Higgs compositeness in principle fits well this situation. That is certainly the case if the large width is into new invisible states. In that case a strong sector characterized by an effective coupling $g_*\sim 3$ among resonances, reminiscent of a theory with large $N\sim 10$, seems like a fair explanation of the data. On the other hand, for  the case in which the width must be accounted by either the decay to $\bar t t$ or to $W_LW_L$/$hh$, parameters have to be somewhat stretched. Needless to say the search for the decay to these other modes will be crucial to decide the plausibility of this option. 

In section~\ref{DM} we find that $S$ decays into Dark Matter can reproduce the large $S$ width
favoured by ATLAS as well as leading to a DM thermal relic abundance compatible with the measured cosmological abundance.

\appendix

\mio{
\begin{figure}[t]
\begin{center}
$$\includegraphics[width=0.55\textwidth]{figs/run1ATLASgammagamma.png}\qquad
\includegraphics[width=0.31\textwidth]{figs/run1CMS-gammagamma.png}
$$
\caption{\em Run 1 $\gamma\gamma$ data. \label{fig:figrun1}}
\end{center}
\end{figure}
}

\footnotesize

\subsubsection*{Acknowledgments}
This work was supported by the ERC grant NEO-NAT and  the MIUR-FIRB grant RBFR12H1MW.
We thank M. Mangano, C. Maiani,  E. del Nobile, A.D. Polosa and G. Salam for useful discussions. The work of JFK was supported in part  by the Slovenian Research Agency.
The work of AP    has been partly supported by 
the Catalan ICREA Academia Program and  grants
 FPA2014-55613-P, 2014-SGR-1450 and  SO-2012-0234.
The work of RR and RT is supported by Swiss National Science Foundation under grants CRSII2-160814 and 200020-150060.

%

%

\end{document}